\documentclass[
reprint, verse,
 amsmath,amssymb,
 aps,prl
]{revtex4-2}
\usepackage{amsmath,amssymb}
\usepackage{fancyvrb}
\usepackage{graphicx}
\usepackage{dcolumn}
\usepackage{bm}
\usepackage{tcolorbox}
\usepackage{hyperref}
\usepackage{makecell}

\makeatletter
\renewcommand\@seccntformat[1]{} 
\makeatother

\usepackage{xr}
\makeatletter

\newcommand*{\addFileDependency}[1]{
\typeout{(#1)}
\@addtofilelist{#1}
\IfFileExists{#1}{}{\typeout{No file #1.}}
}\makeatother

\newcommand*{\myexternaldocument}[1]{%
\externaldocument{#1}%
\addFileDependency{#1.tex}%
\addFileDependency{#1.aux}%
}
\myexternaldocument{supplement}

\makeatletter
\DeclareRobustCommand{\iscirc}{\mathord{\mathpalette\is@circle\relax}}
\newcommand\is@circle[2]{%
  \begingroup
  \sbox\z@{\raisebox{\depth}{$\m@th#1\bigcirc$}}%
  \sbox\tw@{$#1\square$}%
  \resizebox{!}{\ht\tw@}{\usebox{\z@}}%
  \endgroup
}

\begin{document}

\title{Embodied intelligence solves the centipede's dilemma}

\author{Adam Dionne$^{1}$}
\author{Fabio Giardina$^{1}$}
\author{L. Mahadevan$^{1,2}$}

\affiliation{$^{1}$School of Engineering and Applied Sciences, Harvard University, Cambridge, USA.
}
\affiliation{$^{2}$Departments of Physics, and Organismic and Evolutionary Biology
 Harvard University, Cambridge, USA.
}
\date{\today}

\begin{abstract}
Although commonly associated with limbless animals like snakes and fish, multi-legged organisms like centipedes also utilize undulatory locomotion. Whether these undulations are actively reinforced or resisted by the axial musculature remains an open question. We present a dynamical model of centipede locomotion that integrates leg-ground interactions, passive body mechanics, and active lateral musculature. By varying stepping rate, actuation, and body stiffness, we examine how locomotor strategies affect speed and an effective energetic efficiency. Coordination emerges only when body stiffness is tuned to stepping frequency: overly flexible bodies lose synchrony, while overly rigid ones move slowly and inefficiently. This leads to the prediction that centipedes utilize speed dependent active stiffness to maintain this coordination.  Our results suggest that lateral muscles also have a speed dependent function, revealed by optimizing speed and an effective cost, that resists a phase lag between leg touchdowns and body curvature. Together, we find that centipedes actively modulate body mechanics to achieve rapid, efficient locomotion, highlighting how complex control can emerge from embodied physical properties rather than solely from neural computation.
\end{abstract} 

\maketitle

\section{Significance Statement}
Multi-legged animals coordinate dozens of moving parts to walk and run effectively, yet many achieve this without complex central control. Using a minimal neuromechanical model of centipede locomotion, we show that coordination and speed emerge when the body’s stiffness is tuned to the timing of leg-ground contact. Faster gaits require proportionally stiffer bodies, a prediction that implies active stiffness modulation in living animals. 
We also find that axial musculature coordinates legged actuation with the body's curvature, resolving a long-standing debate about their function. 
These results reveal how evolution can exploit body mechanics to enable fast, efficient locomotion and provide design principles for robust, decentralized walking robots.

\section{Introduction}
\label{sec:intro}

Animals routinely coordinate bodies with many degrees of freedom to produce rapid, stable locomotion~\cite{Gray1968}. This task is particularly challenging in organisms with elongated bodies and numerous limbs, where dozens of joints and muscles must act in concert to generate coherent motion. 
Since many of the first land animals had long slender bodies bearing numerous legs~\cite{Manton1965,Jung2018,giardina2021models}, this coordination problem represents an important step in the evolution of terrestrial animals. While such coordination is often attributed to complex neural control, many biological systems instead rely heavily on the physical properties of their bodies and their interaction with the environment. 
In these systems, coordination might well have emerged through embodied intelligence, in which mechanics, actuation, and environmental forces together produce organized behavior without centralized computation.

Multi-legged arthropods provide a striking example of this principle, including the early terrestrial arthropods which were among the first animals to colonize land \cite{Manton1965,Jung2018,giardina2021models}. Fossil and comparative evidence suggests that these organisms achieved increasingly rapid locomotion without dramatic increases in neural complexity \cite{Manton1951,Manton1965}. Instead, locomotor performance correlates with morphological changes such as increasingly elaborate dorsal musculature \cite{Manton1965}. In extant centipedes, locomotion is largely organized locally: each body segment contains paired segmental ganglia controlling the legs and musculature of that segment \cite{SmarandacheWellmann2016,Yasui2019}. These observations suggest that coordination in multi-legged locomotion may arise primarily from tuning the body’s physical properties rather than centralized neural control. Understanding how such organisms achieve rapid and robust locomotion therefore sheds light on an important stage in the evolution of terrestrial movement and provides inspiration for decentralized control strategies in multi-legged robots \cite{full1989,Hoffman2012,Aoi2016,Aoi2023,Chong2022,Chong2023,DooyeolKoh2010,Yasui2017}.

The  robustness of this decentralized organization is highlighted by the fact that centipedes can continue coordinated locomotion even after decapitation \cite{Zareen2016ResponsiveBO}. This observation raises a broader question: how can organisms with many degrees of freedom coordinate their movements without centralized control? The puzzle is captured by ``The Centipede’s Dilemma'', written by Katherine Craster and popularized by Lankester \cite{lankester1889muybridge}:
\begin{verse}A centipede was happy – quite!\\
Until a toad in fun\\
Said, ``Pray, which leg moves after which?''\\
This raised her doubts to such a pitch,\\
She fell exhausted in the ditch\\
Not knowing how to run.\\
\end{verse}
The ditty highlights an apparent paradox: coordinated locomotion involving many limbs seems to require awareness of a complex stepping sequence. Yet biological organisms clearly achieve such coordination without explicit awareness of the process. Understanding how this occurs remains a central problem in the mechanics and control of locomotion.

A persistent obstacle to resolving this puzzle concerns the role of lateral body undulations in multi-legged locomotion \cite{Guo2008,Mahadevan2003}. In limbless animals such as snakes, undulations provide the primary propulsive mechanism \cite{Gray1950}. In organisms such as centipedes, however, propulsion arises from both leg forces and axial bending \cite{Anderson1995,Nirody}. At high speeds centipedes exhibit pronounced lateral undulations \cite{Anderson1995,Manton1951,Kuroda2022}, yet it remains unclear whether these motions are actively generated by axial musculature or passively imposed by leg forces resisted by the trunk. Experimental measurements show muscle activity approximately in phase with joint flexion \cite{Anderson1995}, suggesting active reinforcement, whereas classical morphological arguments propose that fast-running lineages should suppress undulation to avoid energetic losses \cite{Manton1965}. Resolving this apparent contradiction requires a framework that integrates neural actuation, body mechanics, and environmental interaction.

Here we develop a dynamical model of multi-legged locomotion that explicitly couples leg–ground interactions, axial elasticity, and active bending torques. Each body segment interacts with the ground through intermittent leg contacts while being mechanically coupled to its neighbors through elastic and dissipative elements. Active lateral bending is modeled as a traveling wave of internal torque whose phase relative to leg stepping can be varied. Crucially, the model imposes no explicit coordination or feedback control.

Nevertheless, coordinated locomotion, stability, and speed emerge through appropriately tuned passive and active body mechanics. We find that coordination requires the body to be sufficiently stiff to act as a mechanical low-pass filter on contact impulses and actuation so that body undulations remain synchronized with the stepping pattern. As locomotion speed increases, this stiffness must also increase to preserve coordination. Using kinematic data for the giant desert centipede \textit{Scolopendra heros} from Anderson et al. \cite{Anderson1995}, we predict roughly a seven-fold increase in effective intersegmental stiffness across observed running speeds and propose a minimal non-invasive experiment—adapted from classical snake locomotion studies \cite{Gray1950}—to test this prediction.

Finally, by optimizing simultaneously for speed and energetic cost, we show that the functional role of axial musculature depends on operating speed. At low speeds, passive body mechanics can coordinate leg actuation with body curvature such that axial muscle activity is inefficient. At high speeds, active bending contributes directly to propulsion until the limit of coordination is reached. At intermediate speeds, axial musculature primarily enhances coordination by reducing the phase lag between leg touchdown and body curvature. This speed-dependent transition reconciles conflicting experimental interpretations and suggests that axial muscles function as adaptive mechanical elements whose role shifts across locomotor regimes.

Together, these results show that tuning body mechanics with actuation can produce coordination, speed, and efficiency without centralized feedback control. More broadly, the results suggest that evolutionary improvements in locomotor performance likely arose from morphological and material changes rather than increased neural complexity, while also providing design principles for decentralized, high-speed multi-legged robots.

\section{Computational Model}
\begin{figure*}[ht!]
    \centering
    \includegraphics[width = \linewidth]{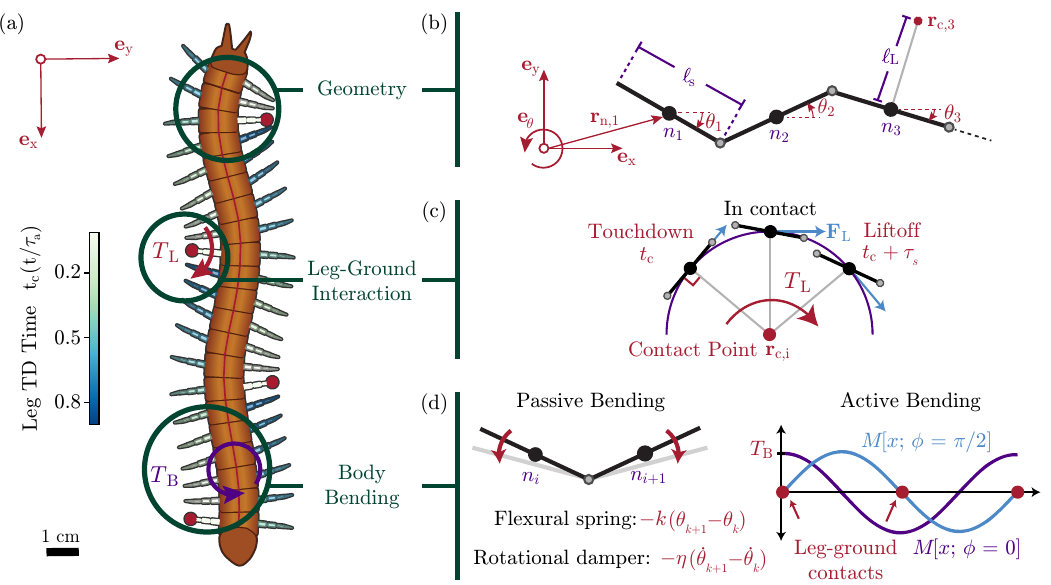}
    \caption{Minimal model of centipede locomotion with legged propulsion and body undulation.
    (\textbf{a}) Schematic of a running centipede showcasing the stepping wave pattern.
    Legs are colored according to their touchdown time $t_c / \tau_a$, where $\tau_a$ is the actuation cycle period. 
    Red dots indicate legs currently in ground contact, which remain in contact for duration $\tau_s$. 
    The schematic represents a 21-segment centipede with $\tau_a/\tau_s = 12$, corresponding to twelve segments spanning between consecutive ground contacts along one side. 
    Scale bar: 2 cm, based on \textit{Scolopendra heros} \cite{Anderson1995}.
    (\textbf{b}) The model's planar geometry. Body segments are modeled as rigid rods. The rod's center is node $n_i$ and has Cartesian position $\textbf{r}_{n,i}$. Segments can rotate in the plane, encoded by the counter-clockwise angle $\theta_i$ relative to horizontal. Two length scales define the geometry: the segment length $\ell_s$ and leg length $\ell_L$. Each body segment has two attached legs, and we denote ground contact's Cartesian position by $\textbf{r}_{c,i}$. 
    (\textbf{c}) Leg-ground interaction. 
    During ground contact (duration $\tau_s$), the attached segment pivots in a circular arc about the contact point due to no-slip constraints (Eq.~\ref{eq:no-slip}) and active leg torque $T_L$ and force $F_L$ (Eq.~\ref{eq:leg_torque}).
    (\textbf{d}) Body bending.
    Passive bending between segments is modeled by a flexural elasticity (spring constant $k$) and a viscous damping (coefficient $\eta$), governed by the inter-segment angle $\sigma_i = \theta_{i+1} - \theta_{i}$ (Eqs.~\ref{eq:flexure} and \ref{eq:damp}). 
    Active bending results from lateral flexor muscles, modeled as a sinusoidal bending moment $M[x,t]$ with amplitude $T_B$ that propagates along the body (Eq.~\ref{eq:bend-moment}). 
    The phase relationship $\phi$ between muscle activation and leg stepping determines whether body undulation assists or resists forward locomotion.} 
    \label{fig:model}
\end{figure*}

\subsection{Geometry and Constraints}
Multilegged organisms, such as centipedes, have bodies composed of many segments, each with a pair of legs. We model each segment as a rigid rod of mass $m$ and length $\ell_s$, with a pair of legs connected to its central node. Without any ground constraints, a chain of $N$ rods in the plane has $N+2$ degrees of freedom associated with an overall translation and a rotation for each rod. 
To express our segmental body's position without redundant degrees of freedom, we use generalized coordinates. 
Let $\mathbf{r}_{n,1} = (x,y)^{\top}$ be the Cartesian position of the posterior segment's central node. 
The position of the $i$-th node, $\mathbf{r}_{n,i}$, is determined by the angle $\theta_i$ of the $i$-th segment relative to the horizontal (Fig. \ref{fig:model}b). 
The system's generalized coordinates are
\begin{align}
\label{eqn:coords}
\textbf{q} = (x, y, \theta_1, \theta_2, \ldots, \theta_N)^T,
\end{align}
where $N$ is the number of segments. 
We can find the Cartesian position of the $i$-th node, $\mathbf{r}_{n,i}$ by
\begin{align}
    \textbf{r}_{n,i} = \textbf{r}_{n,i-1} + \frac{\ell_s}{2} \begin{pmatrix} \cos \theta_{i-1} + \cos \theta_i \\
   \sin \theta_{i-1} + \sin \theta_i 
    \end{pmatrix}.
\end{align}
With this map we can define the translational Jacobian that maps generalized velocity to Cartesian velocity (SI), 
$$
\label{eq:trans_jacob}
\dot{\mathbf{r}}_{n,i} = \mathbf{J}_{T,i} \dot{\mathbf{q}},$$
which we will use to express the active leg force in the generalized coordinate frame (see Eq. \eqref{eq:leg_torque}). 

Leg-ground contact introduces a constraint on the body. 
Let $\mathbf{r}_{c,i}$ be the ground contact point for the $i$-th segment's leg. 
While in contact, the leg's segment pivots in a circular arc about the fixed point $\mathbf{r}_{c,i}$ (Fig. \ref{fig:model}c). 
We enforce this with a no-slip condition acting between the ground contact point and its velocity, 
\begin{align}
\label{eq:no-slip}
\left(\dot{\mathbf{r}}_{n,i} - \dot{\mathbf{r}}_{c,i} \right) \cdot \left(\mathbf{r}_{n,i} - \mathbf{r}_{c,i}\right) = 0.
\end{align}
Using the translational Jacobian and noting that $\dot{\mathbf{r}}_{c,i} = 0$ for a fixed contact point, 
\begin{align}
\underbrace{(\mathbf{r}_{n,i} - \mathbf{r}_{c,i})\mathbf{J}_{T,i}}_{= \mathbf{J}_{c,i}}\, \dot{\mathbf{q}} = 0,
\end{align} where we've defined the contact Jacobian  $\mathbf{J}_{c,i}$ which enforces the no-slip condition, Eq. \eqref{eq:no-slip}, on the generalized velocities $\dot{\mathbf{q}}.$

We then derive our system's equations of motion in generalized coordinates from linear and angular momentum balance, which yields (SI)
\begin{align}
\label{eq:system}
\mathbf{G} \ddot{\mathbf{q}} + \mathbf{h} \dot{\mathbf{q}}  = \boldsymbol{\mathcal{T}}[\mathbf{q},\dot{\mathbf{q}}] + \mathbf{J}_c^T \boldsymbol{\mu},
\end{align}
where $\mathbf{G}$ is the generalized mass matrix, $\mathbf{h}$ is the Coriolis/centrifugal term, $\boldsymbol{\mathcal{T}}[\mathbf{q}, \dot{\mathbf{q}}]$ are the generalized torques and forces, and $\boldsymbol{\mu}$ is the vector of Lagrange multipliers that enforce contact constraints. To close $\boldsymbol{\mathcal{T}}[\mathbf{q}, \dot{\mathbf{q}}]$, we consider bending, both passive and active, along with legged actuation. 

\subsection{Passive Bending}
Passive mechanical interactions between segments are modeled using linear flexural springs and dampers \cite{Yasui2017}, as depicted in Fig. \ref{fig:model}(d). Bending originates at inter-segment joints, and depends on the joint angle $\sigma_i = \theta_{i+1}-\theta_{i}$. The $i$-th segment has an anterior and posterior joint, both of which contribute to the flexural torque
\begin{align}
    \label{eq:flexure}
    S_i = -k(\theta_{i} - \theta_{i-1}) - k(\theta_i - \theta_{i+1}),
\end{align}
where $k$ is the inter-segmental flexural stiffness. Similarly, the rotational damping torque is
\begin{align}
    \label{eq:damp}
    D_i = -\eta(\dot{\theta}_i - \dot{\theta}_{i-1}) - \eta(\dot{\theta}_i - \dot{\theta}_{i+1}),
\end{align}
where $\eta$ is the damping coefficient. These passive torques are collected into matrices that act on the generalized coordinates, $\mathbf{S} = \mathrm{blkdiag}[\mathbf{0}_2, \mathrm{tridiag}[-k,2k,-k]]$ and $\mathbf{D} = \mathrm{blkdiag}[\mathbf{0}_2, \mathrm{tridiag}[-\eta,2\eta,-\eta]]$, to write
\begin{align}
\label{eq:passive_bend}
    \boldsymbol{\mathcal{T}}_{\substack{\text{Passive Bending} }} = \mathbf{S}\mathbf{q} + \mathbf{D}\dot{\mathbf{q}}.
\end{align}

\subsection{Legged Actuation}
Legs in contact with the ground generate a propulsive force $\mathbf{F}_L$ by pushing on the ground, creating a torque $\mathbf{T}_L$ that accelerates the attached segment. For the $i$-th segment, 
\begin{align}
\label{eq:leg_torque}
\mathbf{T}_L = (\mathbf{r}_{n,i} - \mathbf{r}_{c,i}) \times \mathbf{F}_L.
\end{align}
In generalized coordinates, this actuation contributes to both the rotational and translational motion of the segment. Torque acts directly on the $i$-th segment's angular velocity. The translational Jacobian (see Eq. \eqref{eq:trans_jacob}) lets us represent the force as acting on the generalized coordinates. Together, 
\begin{align}
\label{eq:TL}
\boldsymbol{\mathcal{T}}_{\text{Active Legs}} = \mathbf{J}_{R,i}^T \mathbf{T}_L + \mathbf{J}_{T,i}^T \mathbf{F}_L,
\end{align}
throughout a leg's contact. 

Each leg has a touchdown time $t_c$ at which the leg makes ground contact and remains in contact for the step duration $\tau_s$, i.e. until $t = t_c + \tau_s$ (Fig. \ref{fig:model}c). Otherwise, legs are in swing and remain out of contact with the ground throughout the actuation cycle. The contact point $\mathbf{r}_c$ is determined at $t_c$ by the leg's angle of attack $\alpha$. We set $\alpha = \pm \pi/2$ such that legs are orthogonal to their segment at touchdown. 

Leg touchdown times $t_c$ propagate along the body (Fig. \ref{fig:model}a) such that when a leg ends contact the next leg towards the posterior starts contact. This wraps around the body such that when the posterior legs end contact the anterior legs starts contact. Let $t^{i,R}_c$ denote the leg touchdown time for the $i$-th segment's right leg, and $t^{i,L}_c$ denote the $i$-th segment's left leg. We begin the actuation cycle with the posterior segment's right leg starting contact, i.e. $t^{1,R}_c = 0$. Touchdown times are then propagated along the body based on the step duration $\tau_s$. We reduce these time modulo the actuation time period $\tau_a$: 
$$t_c^{i,R} \equiv (1-i) \tau_s \ \   (\mathrm{mod}\  \tau_a).
$$
Inter-segmental legs are offset by half the cycle, such that the left leg makes ground contact at $t_c^{i,L} \equiv  t_c^{i,R} + (\tau_a/2) \ \ (\mathrm{mod}\ \tau_a)$.

The change in contact constraints at each step introduces a discrete impact event, resulting in an instantaneous change in the system's velocity. This velocity update is given by (SI)
\begin{equation}
\label{eq:nonsmooth}
\dot{\mathbf{q}}^+ = \textbf{G}^{-1} \textbf{J}_c^{\top} d\boldsymbol{\mu} + \dot{\mathbf{q}}^-, 
\end{equation}
where $d\boldsymbol{\mu}$ is the constraint impulse.

\subsection{Active Bending}
Active bending moments are generated by the lateral flexor muscles, which are modeled as a traveling wave of torque propagating along the body \cite{Anderson1995} (Fig. \ref{fig:model}d). Letting $x$ be the distance while traversing along each segment's length, the internal bending moment $M_{\text{bend}}$ at position $x$ and time $t$ is
\begin{align}
\label{eq:bend-moment}
M_{\text{bend}}[x,t] = T_B \sin\left[ \frac{2\pi t}{\tau_a} + \frac{2\pi x}{\lambda} + \phi\right],
\end{align}
where $\lambda = \ell_s (\tau_a / \tau_s)$ is the wavelength of the actuation wave, and $\phi$ is the phase lag between the bending wave and the leg stepping pattern. The net torque on the $i$-th segment is the difference in the bending moment across its length,
\begin{align}
\label{eq:TB}
\begin{split}
\mathcal{T}_{\text{Active Bending},i} [t]= \, &M_{\text{bend}}[i\ell_s,t] \\&-M_{\text{bend}}[(i-1)\ell_s,t].\end{split}\end{align}
For the posterior and anterior segment, we use a free boundary condition to determine the bending moment. 

\subsection{Dynamics}
\label{sec:dynamics}
By substituting external forcing from Eqs. \eqref{eq:passive_bend}, \eqref{eq:TB}, and \eqref{eq:TL} into our system constraints from Eq. \eqref{eq:system}
we get our model's dynamical equations,
\begin{align}
\label{eqn:motion}
\begin{split}
\underbrace{\mathbf{G}\ddot{\textbf{q}}}_{\substack{\text{Inertial}\\ \text{Forces}}} + \underbrace{\textbf{h} \dot{\mathbf{q}}}_{\substack{\text{Gyroscopic} \\ \text{Accelerations}}} + \underbrace{\textbf{J}_c \cdot \boldsymbol{\mu}}_{\substack{\text{Leg-Ground} \\ \text{Contacts}}} = \\\underbrace{\textbf{S} \textbf{q}}_{\substack{\text{Flexural} \\ \text{Spring}}} + \underbrace{\textbf{D}  \dot{\textbf{q}}}_{\substack{\text{Rotational} \\ \text{Damper}}} + \underbrace{\boldsymbol{\mathcal{T}}_{\text{Active}}}_{\substack{\text{Leg \& Body} \\ \text{Torques}}}.
\end{split}
\end{align}

In total, we built a rigid body model of multi-legged locomotion with active forces due to legged and body actuation. The model is hybrid, cycling between a continuous contact phase, in which the system's state $\mathbf{q}$ evolves according to Eq. \eqref{eqn:motion}, and discrete leg-ground contact transition events, at which the system's velocities $\dot{\mathbf{q}}$ discontinuously update due to the constraint change via Eq. \eqref{eq:nonsmooth}.

\subsection{Dimensional Analysis and Parameters}
To identify the minimal parameter set controlling locomotion, we find four characteristic time scales by balancing a segment's inertia with the legged actuation torque $T_L$, bending actuation torque $T_B$, and inter-segmental stiffness $k$:
\begin{align}
\tau_{T_L}=\sqrt{\frac{m\ell_s^2+I}{T_L}} \;\; \text{,}\;\;
\tau_{T_B}=\sqrt{\frac{m\ell_s^2+I}{T_B}} \;\; \text{,}\;\;
\tau_{k}=\sqrt{\frac{m\ell_s^2+I}{k}}.
\end{align}
Here $I$ is a segment's moment of inertia, a parameter within the mass matrix $\mathbf{G}$ (SI). 
Coordination is governed by how these intrinsic times compare to the step time $\tau_s$ for which a leg remains in ground contact. We refer to $\tau_s/\tau_{T_L}$ and $\tau_s/\tau_{T_B}$ as the contact--actuation ratios, and $\tau_s/\tau_K$ as the contact--elastic ratio. 

The damping factor $\eta$ is best represented with a damping factor $\zeta$, related by $\eta = \zeta \tau_a/\tau_k.$ We omit the damping factor $\eta$ from the main text's analysis to reduce complexity, motivated by the relative insensitivity of the body's speed to damping compared to the other parameters (SI Fig. 2).

To relate spatial and temporal organization, note that the imposed stepping wave advances by one segment per stance, giving
\begin{align}
\frac{\lambda}{\ell_s} = \frac{\tau_a}{\tau_s},
\end{align}
where $\lambda$ is the actuation wavelength. 

The minimal parameter set used to organize the results can therefore be written as
\begin{align}
\label{eq:params}
\left\{
\frac{\lambda}{\ell_s},\
\frac{\tau_s}{\tau_k},\
\frac{\tau_s}{\tau_{T_L}},\ 
\frac{\tau_s}{\tau_{T_B}},\
\phi
\right\}.
\end{align}

\subsection{Numerical Implementation}
\label{sec:numerics}

\textit{Scolopendra heros} have $N = 21$ segments, and as such $\mathrm{dim}[\mathbf{q}] = 23$. We will investigate the initial conditions $\mathbf{q} = \mathbf{0}$, for which the body begins still and aligned with no flexure. 

The system's parameters, and the ranges explored in our simulations, are summarized in Table \ref{tb:params}. All parameters were non-dimensionalized using the segment mass $m$, segment length $\ell_s$, and the leg actuation period $\tau_a$ throughout implementation. The model was simulated in MATLAB (R2025b), and the smooth dynamics were integrated using the \verb+ode45+ solver, an explicit Runge-Kutte $(4,5)$ solver. The relative and absolute error tolerances were set to $10^{-3}$ and $10^{-6}$ respectively. All code is freely available online \cite{github}.

\begin{table}[ht!]
\renewcommand{\arraystretch}{1.25}
\centering
\begin{tabular}{c l c c}
\hline
\textbf{Symbol} & \textbf{Description} & \textbf{Value(s)} & \textbf{Ref.} \\
\hline
$N$& Number of segments                    & 21 & \cite{Manton1965} \\
$\alpha$        & Leg touchdown angle                     & $\pm \pi/2$ & \\
$I / m \ell_s^2$ & Moment of inertia          & $1/12$ & \\
$\ell_s/\ell_L$ & Segment-leg length ratio & 1.5 & \cite{Manton1965} \\
$\phi$          & Bending-leg phase shift & $[-\pi,\pi]$ & \\
$\lambda/\ell_s = \tau_a/\tau_s$ & Actuation wavelength & $[5,13]$ & \cite{Manton1951, Anderson1995} \\
$\tau_s/\tau_{k}$ & Contact-elastic ratio & $[0,4]$ \\
$\tau_s/\tau_{T_L}$ & Contact-bending actuation ratio & $[0,2]$ \\
$\tau_s/\tau_{T_B}$ & Contact-legged actuation ratio & $[0,2]$ \\
$\zeta$ & Damping factor & $[0,1.5]$ \\ 
\hline
\end{tabular}
\caption{The model's key parameters. Parameters are non-dimensionalized by the segment mass $m$, segment length $\ell_s$, and the actuation time period $\tau_a$.}
\label{tb:params}
\end{table}

\section{Results}

\begin{figure*}
    \centering
    \includegraphics[width=1.0\linewidth]{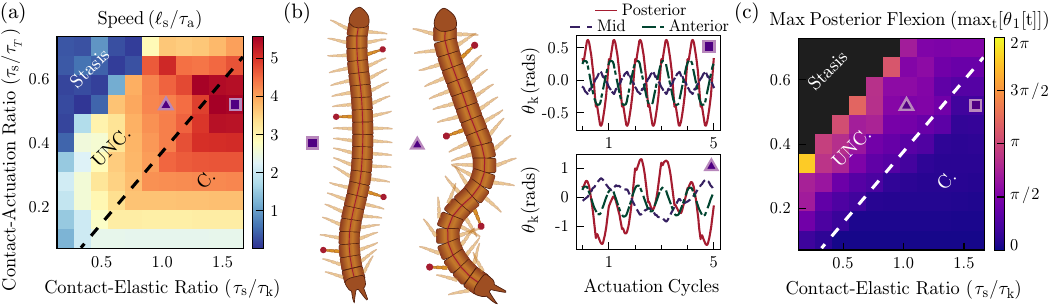}
    \caption{Coordinated locomotion requires a sufficiently high contact-elastic ratio $\tau_s/\tau_k$ and a sufficiently low contact-actuation ratio $\tau_s/\tau_T$. Model solutions obtained by integrating Eqs.~\eqref{eq:nonsmooth}, \eqref{eqn:motion} with initial conditions $\mathbf{q}, \dot{\mathbf{q}} = 0$. 
    Parameter values fixed throughout are found in Table \ref{tb:params}, and the varied parameters used here are 
    $\lambda/\ell_s = 11, \tau_s/\tau_k \in [0.2,1.6], \tau_s/\tau_{T_L}  \in [0.1,0.7], \tau_{T_L}/\tau_{T_B} = 1.36, \phi = 4\pi/5.$
    (\textbf{a}) Phase diagram of the body's center of mass speed $(\ell_s/\tau_a)$ versus contact-elastic ratio $(\tau_s/\tau_k)$ and contact-actuation ratio $(\tau_s/\tau_T)$. 
    Three distinct regimes emerge: uncoordinated motion (UNC.), coordinated locomotion (C.), and stasis.
    The dashed line marks the coordination boundary, determined by the transition from multi-periodic to single-periodic segment oscillations (Fig. S3). 
    In the stasis regime, speed approaches zero as segment flexion increases to the point where leg pivots no longer align to give forward propulsion.
    (\textbf{b}) Comparison of uncoordinated ($\square$) and coordinated ($\triangle$) locomotion showing body configurations and temporal dynamics.
    Time series show segment angles $\theta_i$ for posterior ($i=1$), middle ($i=11$), and anterior ($i=21$) segments over five actuation cycles.
    Coordinated locomotion exhibits regular periodic oscillations, while uncoordinated motion shows irregular behavior.
    The posterior segment consistently exhibits the largest amplitude oscillations.
    (\textbf{c}) Maximum posterior segment flexion $(\mathrm{max}_t[\theta_1[t]])$ across the same parameter space as panel (a).
    The sharp increase in posterior flexion at the coordination boundary indicates that loss of coordination is driven by excessive bending of the posterior segment.}
    \label{fig:coordination}
\end{figure*}

We begin by investigating how the model approaches a steady state. Considering purely legged, purely bending, and mixed actuation strategies, we find that the body's acceleration over time is similar up to scaling across these actuation strategies (Fig. S1). Visualizing the motion (Movie 1), we find that the model's motion agrees with the observed qualitative description of centipede locomotion: lateral bending with the same time period as the leg's stepping pattern, and leg touchdowns temporally localized with maximal segmental flexion. 

To understand how the body flexion coordinates with actuation, we investigate what parameters break this synchrony. Varying the contact-elastic ratio, $\tau_s/\tau_k$, and the contact-actuation ratio, $\tau_s/\tau_T$, reveals three distinct regimes of body organization --- coordinated, uncoordinated, and stasis --- as labeled in Fig. \ref{fig:coordination}a. The body coordinates, resulting in steady state forward locomotion, only when the contact-elastic ratio ($\tau_s/\tau_k$) is sufficiently stiff, and the contact-actuation ratio ($\tau_s/\tau_T$) is sufficiently small.
We identify a solution as uncoordinated if the segment angles $\theta_i$ are not $\tau_a$-periodic, and in stasis when the effective propulsive force makes a discontinuous jump (Fig. S3, Eq. \eqref{eq:cost}). 
If the body is too flexible or the actuation torque too large, the traveling wave of leg touch downs fails to synchronize such that during touchdown leg actuation is no longer mostly parallel to travel, resulting in a loss of effective propulsion as the body and actuation synchrony breaks down (Fig. \ref{fig:coordination}b--c, Movie 1). 

In an extreme region of high torque and low stiffness, the body enters a stasis regime where segments are highly flexed, the state evolves chaotically, and the effective propulsive force (see Eq. \eqref{eq:cost}) has a jump increase (Fig. S3). The regime is non-physical, as steric interactions between legs and segments become essential. Overall, Fig. \ref{fig:coordination} suggests that a key function of the centipede's passive body stiffness is to mechanically couple the segments, acting as a low-pass filter on contact impulses and actuation such that the lowest frequency mode corresponding to the actuation wavelength is dominantly excited. 

\begin{figure*}
    \centering
    \includegraphics[width = 1\linewidth]{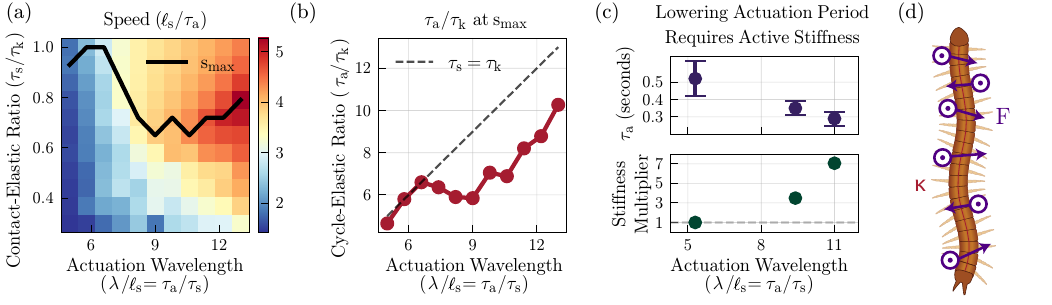}
  \caption{Running speeds require longer stepping wavelengths and proportionally stiffer bodies, with optimal stiffness scaling as $\tau_k \sim \tau_s$.
  Parameter values fixed throughout are found in Table \ref{tb:params}, and the varied parameters used here are
  $\lambda/\ell_s \in [5,13], \tau_s/\tau_k \in [0.3,1], \tau_s/\tau_{T_L} = 0.3, \tau_s/\tau_{T_B} = 0.3, \phi = \pi/2.$
    (\textbf{a}) Speed landscape as a function of actuation wavelength $\lambda/\ell_s = \tau_a/\tau_s$ and contact-elastic ratio $\tau_s/\tau_k$. 
    The black curve traces the maximum achievable speed $s_{\text{max}}$ for each wavelength, revealing an optimal contact-elastic ratio that varies with gait.
    (\textbf{b}) Optimal relationship between elasticity and actuation wavelength: the value of $\tau_a/\tau_k$ that maximizes speed for each wavelength.
    The dashed line ($y = x$) represents the condition $\tau_s = \tau_k$, showing that optimal speed occurs when the contact timescale $\tau_s$ approximately matches the body's elastic response time $\tau_k$.
    (\textbf{c}) Optimal speed scaling suggests active stiffness that increases with speed. 
    Top panel shows experimental results using \textit{Scolopendra heros} data from Anderson et al. \cite{Anderson1995}, which demonstrate a decreasing actuation period $\tau_a$ as speed increases. 
    Bottom panel shows the stiffness multiplier required at each speed to achieve the optimal $\tau_a/\tau_k$ relationship predicted in panel (b) given the measured $\tau_a$ from the top panel.
    This analysis predicts that centipedes must increase body stiffness by up to 7-fold to achieve their highest experimental speeds.
    (\textbf{d}) Inspired by Gray and Lismann \cite{Gray1950}, we propose an experimental test using pendulum force measurements to directly quantify effective inter-segment stiffness in order to test our model's prediction (see Discussion). 
    }
    \label{fig:speed}
\end{figure*}

As shown in Fig. \ref{fig:speed}a, for any given stepping wavelength, there is an optimal body stiffness that maximizes forward speed. This optimal stiffness corresponds to a regime where the contact timescale ($\tau_s$) is closely matched to the body's elastic response timescale ($\tau_k$) (Fig. \ref{fig:speed}b).
Experimental data for the centipede \textit{Scolopendra heros} show that as the animal runs faster, its leg actuation period, $\tau_a$, decreases (Fig. \ref{fig:speed}c, top panel). Since higher speeds are associated with longer wavelengths ($\tau_a/\tau_s$), a decrease in $\tau_a$ implies a decrease in $\tau_s$. As such, for the step-stiffness ratio $\tau_s/\tau_k$ to stay near the optimum curve in Fig. \ref{fig:speed}a, it must be that $\tau_k$ decreases with the actuation wavelength. In turn the inter-segmental stiffness $k$ must increase (since $\tau_k \propto 1/\sqrt{k}$). This implies centipedes actively stiffen their bodies to move faster, which is feasible by isometric contraction of dorsal lateral muscles \cite{Manton1965}. 
A potential benefit to active stiffness is that centipedes utilize their soft body to maneuver between gaps, soil, and bark \cite{Manton1965}. 
Our model predicts a nearly seven-fold increase in body stiffness to achieve the highest observed running speeds (Fig. \ref{fig:speed}c, bottom panel).

Finally, we investigate the trade-offs between a locomotory pattern's speed and efficiency. We approach this by measuring the effective propulsive force required for one steady state cycle, measured as the positive work done by legged and bending actuators divided by the distance traveled: 
\begin{align}
\label{eq:cost}
\frac{\int_0^{\tau_a} (\mathbf{P}^+_{\text{bend}}[t] + \mathbf{P}^+_{\text{leg}}[t]) \, dt }{s_{\text{com}} \tau_a} \equiv \frac{W^+}{d},
\end{align}
where the power due to actuators is given by $\dot{\mathbf{q}}^{\top} \mathcal{T}_{\text{Active}}$. 
This quantity represents an effective minimal requisite propulsive force for a given steady state solution, measured in units of the segmental force scale $m_s\ell_s/\tau_a^2$, which we consider the cost for a given locomotor strategy. We perform a multi-objective optimization using a genetic algorithm to find the Pareto-optimal locomotion strategies that maximize speed while minimizing the cost, i.e. the effective force (SI). We allow the system's non-dimensional parameters, given in Eq. \eqref{eq:params}, to vary in this optimization. 

\begin{figure*}
    \centering
    \includegraphics[width=1.0\linewidth]{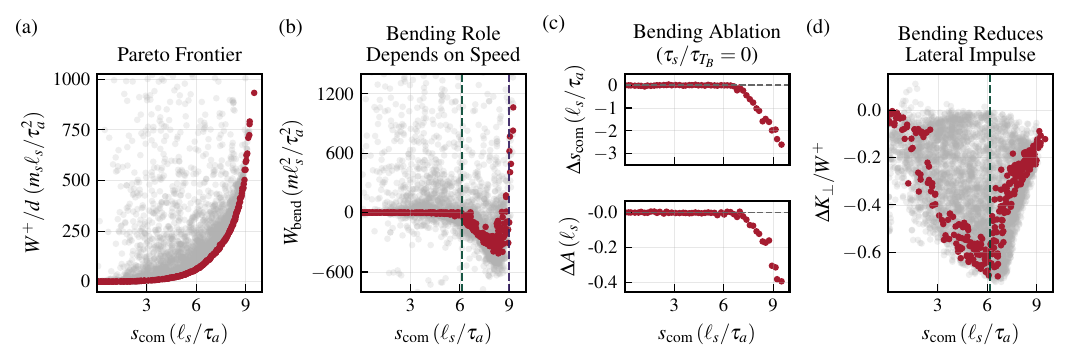}
       \caption{
    Pareto-optimal locomotion strategies reveal speed-dependent transitions from leg dominated to bending dominated locomotion.
(\textbf{a})  Speed-cost trade-offs across parameter space $\{\lambda/\ell_s, \tau_s/\tau_k, \tau_s/\tau_{T_L}, \tau_s/\tau_{T_B}, \phi\}$ with fixed $\eta = 1.25 \tau_a / \tau_k$. The parameter space is adaptively sampled using a genetic algorithm (SI). 
(\textbf{b}) Net bending work $W_{\text{bend}}$ over a steady state cycle along the Pareto frontier exhibits speed-dependent sign changes. For $s_{\text{com}} < 6$, to the left of the dashed green line, the work is small in magnitude and can be resistive or assistive. For $6 < s_{\text{com}} < 9$, between the green and purple dashed lines, bending work is negative and as such in aggregate resists the body's shape change. For top speeds $s_{\text{com}} > 9$, to the right of the purple line, bending work is positive and in net assists the body's shape change. 
(\textbf{c}) By ablating active bending, i.e. setting $\tau_s/\tau_{T_B}=0$, for parameter values along the frontier, we find a decrease in the center of mass speed and lateral undulatory amplitude. As such, although active bending does negative work for $s_{\text{com}} > 6$, it does not resist undulations but rather increases them. 
(\textbf{d}) The normalized impulse energy loss over a cycle lateral to the body's motion, $\Delta K_{\perp}/W^+$, increases in magnitude until $s_{\text{com}} \approx 6$ to oppose $\approx 60\%$ of the actuator's positive work $W^+$. As active bending begins resisting undulations, this proportion decreases at higher speeds.}
    
    \label{fig:pareto}
\end{figure*}

The resulting Pareto frontier, shown in Fig. \ref{fig:pareto}a, reveals the optimal trade-off between speed and cost. Along this front, we find three distinct locomotion strategies deployed that differ by the functional role of active bending: passive coordination, active coordination, and active propulsion. 
These regimes correspond to low, mid, and high speeds as marked by the green and purple dashed lines in Fig. \ref{fig:pareto} at $s_{\text{com}} = 6 \ \ell_s/\tau_a$ and $s_{\text{com}} = 9 \ \ell_s/\tau_a$. 

For passively coordinated locomotion, active bending is not utilized. This is supported by the motion's insensitivity to active bending within the regime: the bending work $W_{\text{bend}}$ is near zero, and bending ablation does not change the body's speed or the body's lateral undulation amplitude (Fig. \ref{fig:pareto}b-c). Lastly, the speed to cost Pareto frontier with bending ablation finds no difference in cost within this regime, but higher costs for higher speeds (Fig. S5).

For actively coordinated locomotion, the joint's lateral musculature works to synchronize leg touchdowns with maximal joint flexure such that legged actuation is delivered in the direction of travel. To understand active coordination, we define two related phase shifts that measure coordination. The first, $\Psi_{\text{leg}}$, is the phase between maximal joint flexure and the leg touchdown time, while the second, $\Psi_{\text{bend}}$, is the phase between maximal joint flexure and the zero of active bending (Eqs. S13--S14). Ideally, $\Psi_{\text{leg}} = 0$ such that a segment's pivot about a leg touchdown is maximally parallel to the direction of travel. Similarly, when $\Psi_{\text{bend}} = 0$ active bending will do purely positive work and assist motion throughout the entirely of the actuation cycle. However, we find a negative phase shift $\Psi_{\text{leg}}$ that steadily increases in magnitude during the passive coordination regime until peaking at $\Psi_{\text{leg}} = -0.31 \pi$ (Fig. S7). The energetic inefficiency is quantified by Fig. \ref{fig:pareto}d, as the energy dissipated due to leg-touchdown impulses perpendicular to the motion per positive actuator work reaches $\Delta K_{\perp} / W^+ = -0.6$. 

To reach higher speeds, bending works to actively coordinate the body. The functional importance of bending in this regime is supported by multiple observations: the contact-bending actuation ratio $\tau_s/\tau_{T_B}$ transitions from a gradual increase along the frontier to a large slope that outpaces the contact-legged actuation ratio (Fig. S4), $W_{\text{bend}}$ becomes negative, bending ablation reduces speed and lateral undulation amplitude, and $\Delta K_{\perp} / W^+$ begins decreasing in magnitude (Fig. \ref{fig:pareto}b--d). Crucially, the phase shift $\Psi_{\text{leg}}$ also begins to decrease in magnitude. This motivates our understanding of the role of active bending in this regime as coordinating the locomotion. We find that the power dissipated due to bending for each joint is temporally localized to the anterior and posterior leg touchdowns. For each joint, bending does positive work at the start of touchdown until the joint reaches maximal flexure. Throughout the remainder of the touchdown bending resists the pivot. As such, bending actively coordinates the leg-touchdowns with the body's curvature by accelerating segments towards peak curvature at the start of touchdown such that legs can deliver efficient propulsion (Fig. S7a,d). 

For top speeds, we reach the active propulsion regime. Here $W_{\text{bend}}$ and $\Psi_{\text{leg}}$ both change sign to positive (Fig. \ref{fig:pareto}b, Fig. S7a), and as such bending contributes to propulsion. Concurrently, lateral undulations increase in amplitude, including the four posterior segments which exhibit the largest lateral displacement (Fig. S7c, S9d). Active propulsion exists close to the limit of coordination between the body and legs. The regime begins at $s_{\text{com}} = 9 \ \ell_s/\tau_a$ and ends at the maximum speed $s_{\text{com}} = 9.51 \ \ell_s/\tau_a$, after which the body cannot increase its center of mass speed. The peak speed is concurrent with peak lateral amplitude which plateaus for higher actuation torque (Fig. S9a). As such, the body likely reaches a kinematic limit to the lateral amplitude and further driving torque channels energy into shape change rather than propulsion.

Given a centipede's center of mass $s_{\text{com}}$, we cannot use Fig. \ref{fig:pareto} to determine whether the centipede is utilizing passive coordination, active coordination, or active propulsion. This is because animals traveling at a given speed will trade off cost for stability. 
For our model, stability is primarily dependent on the actuation wavelength $\lambda/\ell_s$ since the average number of legs in contact with the ground throughout an actuation cycle is $2N \ell_s/\lambda$.
As such, we can characterize the stability trade off by considering the speed versus cost Pareto frontier for fixed actuation wavelength $\lambda/\ell_s$. 
For three trials with fixed $\lambda/\ell_s$, we find that the resulting Pareto frontier reaches a smaller maximum speed but are similar up to scaling with the frontier found in Fig. \ref{fig:pareto}. Each trail exhibits the same transitions between passive coordination, active coordination, and active propulsion (Fig. S6). 
As such, active coordination is likely the dominant strategy utilized at most speeds while passive coordination and active propulsion only occur for the slowest or fastest observed speeds.

\section{Discussion}
\label{sec:discussion}

The centipede's dilemma addresses a natural question underlying the dynamics (and evolution) of locomotory gaits in elongated multi-legged organisms: how can one coordinate dozens of legs, body segments, and muscles to move fast? Our work begins to answer the dilemma by showing how coordination emerges without any feedback, via tuned body mechanics, when inter-segment stiffness is strong enough to act as a low-pass filter on the high frequency contact impulses and actuation. Then, fast movement emerges when the contact time for a step matches the inertial timescale set by the stiffness. In this view, the body undulation acts as a low-dimensional template that organizes leg movements through mechanical coupling. Because many legs interact with the ground simultaneously, small changes in body wave amplitude or phase propagate along the body and generate coherent changes in propulsion and turning.

Our theory predicts that centipedes utilize the dorsal oblique musculature to actively increase inter-segment stiffness as speed increases. This model prediction can be validated with a minimal non-invasive experiment to directly quantify the effective body stiffness of a centipede at different speeds (Figure \ref{fig:speed}d). Seventy five years ago, Gray \& Lissmann \cite{Gray1950} used pendulum-based force measurements
 to study snake locomotion  by having the animal move through an array of cylindrical pendula (Fig.~\ref{fig:speed}d). One can study a centipede's body mechanics at varying speed to measure how an undulating body exerts lateral forces against the pendula, causing measurable deflections. The effective inter-segment stiffness 
\begin{align}
    k_{eff} = \frac{F \cdot r}{\Delta \theta}
\end{align}
can then be calculated from the pendulum deflection force $F$, the lever arm $r$, and the measured angular deflection $\Delta \theta$ between adjacent body segments. If centipedes actively increase body stiffness to achieve coordinated high-speed locomotion, as we predict, this experiment would measure an increase in $k_{eff}$ as speed increases. Complementing this, invasive experiments associated with stiffening the centipede's backbone using adjustable joints or neuro-toxic drugs that induce muscle contraction like botulinum toxin (Botox), can probe the stiffness dependence on speed as characterized in Figs. \ref{fig:coordination},\ref{fig:speed}. 

Our combined framework provides a window into the locomotor dynamics of multi-legged organisms, and captures salient features of centipede locomotion and makes testable predictions about how coordination arises and propulsion is optimized. These results also suggest a resolution of the conflict between Manton \cite{Manton1965} and Anderson et al. \cite{Anderson1995} regarding the origin of lateral undulations. We propose that the axial musculature of centipedes does not purely resist or purely assist undulations, but rather deploys a speed dependent strategy driven by the need for coordination. At the lowest speed, passive coordination is sufficient and axial musculature is not necessary. At high speeds, musculature assists locomotion while pushing to the limit of coordination. For most intermediary speeds, the axial musculature yields active coordination between the legged actuation and the body's curvature by reducing the phase lag $\Psi_{\text{leg}}$. While the bending work does resist the motion, the work actually increases the lateral undulations due to the increased coordination. 

More broadly, our work highlights the importance of the dynamic coupling between the brain, body, and environment  in developing and evolving robust locomotion. Indeed, recent observations of centipede locomotion in cluttered environments suggest that body undulations and leg coordination are adaptable across a wide range of speeds and terrains~\cite{PierceEtAl2026Centipede}. These suggest that locomotor coordination remains robust even when individual leg contacts are perturbed, consistent with the idea that body dynamics provide a global organizing signal. This naturally has implications for  multi-legged locomotion  in biomimetic robotics, where undulatory robots could have advantages traversing rough terrain \cite{Aoi2016,Chong2023}. Recent work~\cite{Flores2025Steering} shows that steering in centipede-inspired robots can be achieved by superimposing a secondary traveling wave on the primary lateral body undulation, and reinforces the idea that body waves provide an effective low-dimensional control template for navigation in complex environments.  

An important question for future work is whether the present framework can reproduce some of the behavioral transitions observed in animals navigating cluttered environments. In particular, changes in body stiffness, wave amplitude, or phase relationships could potentially generate the shifts between normal gait patterns and the collision-induced postures reported in experiments. Understanding how such transitions emerge from body–environment interactions may help unify biological observations with robotic mimics of multi-legged locomotion.

\section{Acknowledgments}
We thank the National Science Foundation Graduate Research Fellowship Program under Grant No. 2140743 (AD), and the Simons Foundation and the Henri Seydoux Fund for partial financial support. Any opinions, findings, and conclusions or recommendations expressed in this material are those of the author(s) and do not necessarily reflect the views of the National Science Foundation. 

\bibliography{refs}

\clearpage

\newcommand{\beginsupplement}{%
        \setcounter{table}{0}
        \renewcommand{\thetable}{S\arabic{table}}%
        \setcounter{figure}{0}
        \renewcommand{\thefigure}{S\arabic{figure}}%
        \setcounter{equation}{0}
        \def\theequation{S\arabic{equation}}

        \setcounter{section}{0}
        \def\thesection{S\arabic{section}}

        \setcounter{subsection}{0}
        \def\thesubsection{\arabic{subsection}}
     }

\onecolumngrid
\beginsupplement

\begin{center}
{\Large \textbf{Supporting Information for \\``Embodied intelligence solves the centipede's dilemma"}}
\end{center}
\tableofcontents

\section{Supplementary Model Details}
\label{SIsec:model}
Here we describe model details not included in the main text.
We start with the inertial mechanics given by the projected Newton Euler equations \cite{ETH}, then consider constraints due to leg ground contacts, and finally compute impulse updates due to constraint switching.

\subsection{Inertial Equations of Motion}
\label{SIsec:EOM}

Our model's state is the centipede segments' positions and velocities. We represent this with the generalized position vector \textbf{q} and velocity vector \textbf{u}, where the first segment is parameterized by Cartesian coordinates while the remaining are parameterized by their angle $\theta_k$:
\begin{align*}
\textbf{q}&= (x, y, \theta_1, \theta_2, \ldots, \theta_n)^T, \\
\dot{\textbf{q}} &= (\dot{x}, \dot{y}, \dot{\theta}_1, \dot{\theta}_2, \ldots, \dot{\theta}_n)^T.
\end{align*}
The $k$-th segment is at the Cartesian position $\mathbf{r}_{n,i}$, and can be determined given the previous segment $\mathbf{r}_{n,i-1}$, finding the connecting joint's position using $\theta_{i-1}$, and finally $\mathbf{r}_{n,i}$ using $\theta_i$:
\begin{align}
    \label{SIeq:rk}
    \textbf{r}_{n,i} = \textbf{r}_{n,i-1} + \frac{\ell_s}{2} \begin{pmatrix} \cos \theta_{i-1} + \cos \theta_i \\
   \sin \theta_{i-1} + \sin \theta_i 
    \end{pmatrix}.
\end{align}
Using this we can define translational Jacobian $\mathbf{J}_{T,i}$ which transforms a segment's velocity in generalized coordinates to Cartesian,
$$
\dot{\mathbf{r}}_{n,i} = \mathbf{J}_{T,i} \dot{\mathbf{q}},$$
For the first segment,
\begin{align}
\mathbf{J}_1 = \begin{bmatrix} 1 & 0 & 0 & \cdots \\ 0 & 1 & 0 & \cdots \end{bmatrix}.
\end{align}
while for segment $\mathbf{J}_{T,i}$ we differentiate Equation \eqref{SIeq:rk} to find
\begin{align}
\boldsymbol{J}_{T,i} = \begin{bmatrix} \cdots & 0 & -\frac{\ell_{s}}{2} \sin \theta_{i-1} & -\frac{\ell_s}{2} \sin \theta_i & 0 & \cdots \\ \cdots & 0 & \frac{\ell_{s}}{2} \cos \theta_{i-1} & \frac{\ell_s}{2} \cos \theta_i & 0 & \cdots \end{bmatrix} + \boldsymbol{J}_{T,i-1}.
\end{align}
We also define a rotational Jacobian, which selects $\dot{\theta}_i$, 
$$
\mathbf{J}_{R,i} = \mathbf{e}_{i+2}$$
where $\mathbf{e}_j$ is the $j$-th standard basis vector. 

We can now write down the inertial components of our mechanics, conserving linear momentum $\mathbf{p}_i$ and angular momentum $\mathbf{N}_i$, in generalized coordinates. Let $m_s, I$ be a segment's mass and moment of inertia, and let $\mathbf{J}_{R,i}$ be the rotation Jacobian. Then, 
$$\sum_{i=1} \left( \mathbf{J}^{\top}_{T,i} \dot{\mathbf{p}}_i + \mathbf{J}^{\top}_{R,i} \dot{\mathbf{N}}_i \right) = \sum_i \left( m_i\mathbf{J}^{\top}_{T,i} \ddot{\mathbf{r}}_{n,i} + I_i \mathbf{J}^{\top}_{R,i} \frac{d}{dt}\left[\mathbf{J}_{R,i}\dot{\mathbf{q}}\right]\right)
$$
Assuming each segment has the same mass $m_s$ and length $\ell_s$, and as such each has the same moment of inertia $I$, we get
\begin{align*}
&= m \left( \boldsymbol{J}_{T,1}^T \ddot{\boldsymbol{r}}_{n,1} + \boldsymbol{J}_{T,2}^T \ddot{\boldsymbol{r}}_{n,2} + \cdots \right) + I \left( \boldsymbol{J}_{R,1}^T \frac{d}{dt} \left( \boldsymbol{J}_{R,1} \dot{\boldsymbol{q}} \right) + \boldsymbol{J}_{R,2}^T \frac{d}{dt} \left( \boldsymbol{J}_{R,2} \dot{\boldsymbol{q}} \right) + \cdots \right) \\
&= m \left( \boldsymbol{J}_{T,1}^T \boldsymbol{J}_{T,1} + \boldsymbol{J}_{T,2}^T \boldsymbol{J}_{T,2} + \cdots \right) \ddot{\boldsymbol{q}} + I \left( \boldsymbol{J}_{R,1}^T \boldsymbol{J}_{R,1} + \boldsymbol{J}_{R,2}^T \boldsymbol{J}_{R,2} + \cdots \right) \ddot{\boldsymbol{q}}  + m \left( \boldsymbol{J}_{T,1}^T \dot{\boldsymbol{J}}_{T,1} + \boldsymbol{J}_{T,2}^T \dot{\boldsymbol{J}}_{T,2} + \cdots \right) \dot{\boldsymbol{q}}, \\ 
&= \textbf{G}\ddot{\textbf{q}} + \textbf{h}\dot{\mathbf{q}}.
\end{align*}
where have define the mass matrix $\textbf{G}$ and the gyroscopic accelerations $\textbf{h}$: 
\begin{align}
    \mathbf{G} &= m\sum_i \mathbf{J}^{\top}_{T,i} \mathbf{J}_{T,i} + I \sum_i \mathbf{J}^{\top}_{R,i} \mathbf{J}_{R,i}, \\
    \mathbf{h}& = m \sum_i \mathbf{J}^{\top}_{T,i} \dot{\mathbf{J}}_{T,i}.
\end{align}

\subsection{Constraints due to Leg Ground Contacts}
\label{SIsec:constraints}
\subsubsection{Computing Lagrange Multipliers}
Consider a leg attached to the segment at position $\mathbf{r}_n$ in contact with the ground at the position $\mathbf{r}_c$. Then let $\textbf{r}_{cn} = \textbf{r}_n - \textbf{r}_c$ denote the vector from the segment to the body. The condition for no-slip contact at point $c$ of a leg at segment $n$ with the ground is
\begin{align*}
    \dot{\textbf{r}}_{cn}^T \textbf{r}_{cn} &= 0, \\
    (\dot{\textbf{r}}_n - \dot{\textbf{r}}_c)^T \textbf{r}_{cn} &= 0, \\
    (\textbf{J}_n  \dot{\mathbf{q}})^T \textbf{r}_{cn} &= 0, \\
    (\textbf{r}_{cn})^T \textbf{J}_n  \dot{\textbf{q}} &= 0,
\end{align*}
where $\textbf{J}_n$ is the Jacobian that maps from generalized coordinate to Cartesian coordinate velocity. We then define the contact Jacobian as
\begin{equation}
\textbf{J}_c = \textbf{r}_{cn}^T \textbf{J}_n.
\end{equation}
To compare our contact constraint to our equations of motion we find $\ddot{\mathbf{q}}$ by differentiating,
\begin{align*}
\dot{\textbf{r}}_{cn}^T \textbf{J}_n  \dot{\textbf{q}} + \textbf{r}_{cn}^T \dot{\textbf{J}}_n  \dot{\textbf{q}} + \textbf{r}_{cn}^T \textbf{J}_n \ddot{\textbf{q}} &= 0, \\
\textbf{J}_c \ddot{\textbf{q}} + \dot{\textbf{q}}^T \textbf{J}_n^T \textbf{J}_n \dot{\textbf{q}} + \textbf{r}_{cn}^T \dot{\textbf{J}}_n \dot{\textbf{q}}&= 0.
\end{align*}
Then if we define 
\begin{equation}
\boldsymbol{\xi} = \dot{\textbf{q}}^T \textbf{J}_n^T \textbf{J}_n \dot{\textbf{q}} + \textbf{r}_{cn}^T \dot{\textbf{J}}_n \dot{\textbf{q}}
\end{equation}
and
\begin{align}
    \boldsymbol{\Gamma} = \boldsymbol{\mathcal{T}}_{\text{Active}} + \textbf{S} \textbf{q} + \textbf{D} \dot{\mathbf{q}} - \textbf{h}\dot{\mathbf{q}},
\end{align}
by comparing our equations of motion and our constraint condition we get 
\begin{equation}
\textbf{J}_c \ddot{\textbf{q}} + \boldsymbol{\xi} = \textbf{J}_c \textbf{G}^{-1} (\boldsymbol{\Gamma} + \textbf{J}_c^T \boldsymbol{\mu}) + \boldsymbol{\xi} = 0.
\end{equation}
With this we can solve for the Lagrange multipliers $\boldsymbol{\mu}$,
\begin{equation}
\boldsymbol{\mu} = -(\textbf{J}_c \textbf{G}^{-1} \textbf{J}_c^T)^{-1} (\textbf{J}_c \textbf{G}^{-1} \boldsymbol{\Gamma} + \boldsymbol{\xi}).
\end{equation}
\subsubsection{Impulse Updates}
When constraints change, we get an impulse update in the generalized velocities from $\dot{\mathbf{q}}^-$ to $\mathbf{q}^{+}$. The new constraint satisfies
$$\mathbf{J}_c \dot{\mathbf{q}}^+ = 0. 
$$
Balancing the change in momentum yields
$$
\mathbf{G}(\dot{\mathbf{q}}^{+} - \dot{\mathbf{q}}^{-}) = \mathbf{J}_c^{\top} d \boldsymbol{\mu},$$
where $d\boldsymbol{\mu}$ are the impulses. By combining these equations, 
$$\mathbf{J}_c (\dot{\mathbf{q}}^{-} + \mathbf{G}^{-1} \mathbf{J}_c^{\top} d \boldsymbol{\mu}) = 0.$$ Solving for the impulses we get
$$d\boldsymbol{\mu} = -(\mathbf{J}_c \mathbf{G}^{-1} \mathbf{J}^{\top}_c)^{-1} \mathbf{J}_c\dot{\mathbf{q}}^{-}.$$
All together, we can now solve for the impulse update in the generalized velocities as
$$\dot{\mathbf{q}}^{+} = \dot{\mathbf{q}}^{-} - \mathbf{G}^{-1}\mathbf{J}_c^{\top} (\mathbf{J}_c \mathbf{G}^{-1} \mathbf{J}^{\top}_c)^{-1} \mathbf{J}_c\dot{\mathbf{q}}^{-}.$$

\section{Supplementary Results}

\subsection{Characterizing the Model's Acceleration and Sensitivity to Damping}
\begin{figure}
    \centering
    \includegraphics{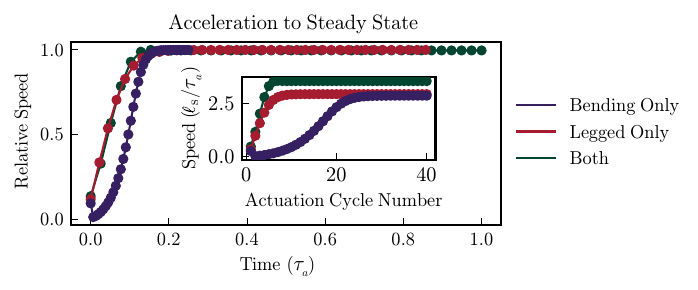}
    \caption{Acceleration to steady state is approximately self-similar under scaling, with at most 30 actuation cycles required for acceleration to reach zero. We simulate locomotion for bending actuation only, legged actuation only, and both. Plotting each solution's speed over time, we find that speed over time is self-similar for the different propulsion strategies. This is visible in the main plot, within which each curve is plotted such that the curves all obtain the same maximum and reach 95\% of that maximum at the same time. The inset displays the unscaled curves. Parameters used: $\lambda/\ell_s = 8, \tau_s/\tau_k = 1.5, \tau_s/\tau_{T_L} = 0.25, \tau_s/\tau_{T_B} = 0.55, \phi = \pi/2, \zeta = 1.25.$}
    \label{SIfig:acceleration}
\end{figure}
We first explore how the model accelerates to its steady state in Figure \ref{SIfig:acceleration}. We find that it takes approximately $30$ full body actuation cycles to reach a steady state, and that the approach to this state is self-similar under scaling. As such, for our results we run all simulations 50 body actuation cycles such that model solutions will reach their steady state. 

\begin{figure}
    \centering
    \includegraphics{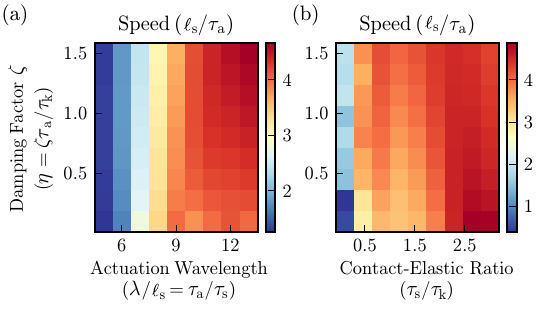}
    \caption{We fix the damping factor $\zeta$ throughout our work to limit complexity since speed is relatively inelastic to changes in $\zeta$. (\textbf{a}) A parameter sweep for the damping factor $\zeta$, where $\eta = \zeta \tau_a/\tau_k$, over the actuation wavelength $\lambda/\ell_s$. (\textbf{b}) A sweep over $\zeta$ and the contact-elastic ratio $\tau_s/\tau_k$. For both, the damping factor $\zeta$ has comparatively little impact on the speed . Parameters used: $\lambda/\ell_s \in [5,13], \tau_s/\tau_k = 2.0, \tau_s/\tau_{T_L} = 0.3, \tau_s/\tau_{T_B} = 0.3, \phi = \pi/2, \zeta \in [0.1,1.5]$.}
    \label{SIfig:damping}
\end{figure}
\label{SIsec:results}

Fig. \ref{SIfig:damping} motivates our de-prioritization of the damping factor $\zeta$, as it is fixed to $\zeta = 1.25$ throughout the main text. Sweeping over $\zeta \in [0.1,1.5]$ while varying both the actuation wavelength and contact-elastic ratio finds that the damping factor's impact on the speed is relatively marginal. As such, we tighten our scope by fixing the damping factor.

 \subsection{Quantifying Coordination}
\begin{figure}
    \centering
    \includegraphics[width=1.0\linewidth]{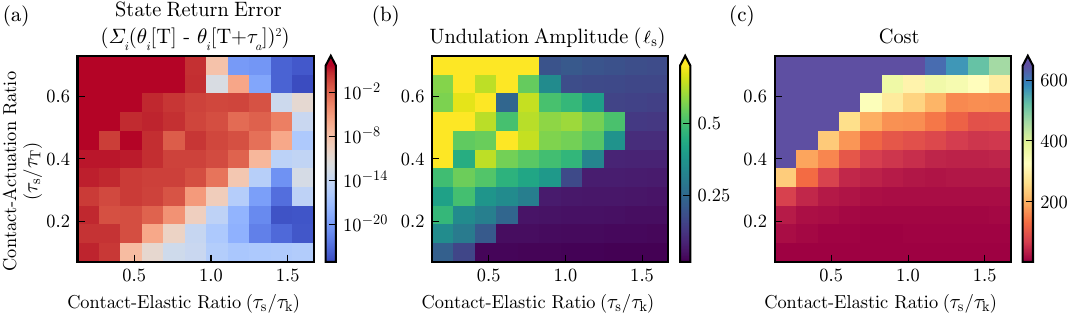}
    \caption{
The metrics used to determine the three regimes of locomotion: coordinated, uncoordinated, and stasis.
  (\textbf{a}) State return error over an actuation cycle, see Eq. \ref{SIeq:return}, as we vary contact-actuation ratio and contact-elastic ratio. Here $T$ is the time at which steady state is reached, i.e. the end of the simulation.
  Low error values (blue) indicate single-periodic motion characteristic of coordinated locomotion, while high error values (red) correspond to multi-periodic or chaotic dynamics in the uncoordinated regime.
  The sharp transition boundary matches the coordination threshold identified in Figure 2 of the main text.
  (\textbf{b}) Undulation amplitude  across parameter space.
  The amplitude displays a jump increase across the coordination boundary.
  (\textbf{c}) The effective cost, defined in the main text as the positive propulsive work over the distance traveled in an actuation cycle $W^+/d$, varies smoothly across the coordination boundary. Yet, decreasing contact-elastic ratio or increasing contact-actuation ratio reveals a discrete jump in cost corresponding to the transition between no coordination and the stasis regime.
  Parameter values match Figure 2 of the main text:$ \lambda/\ell_s = 11, \tau_s/\tau_k \in [0.2,1.6], \tau_s/\tau_{T_L}  \in [0.1,0.7], \tau_{T_L}/\tau_{T_B} = 1.36, \phi = 4\pi/5. $}
  \label{SIfig:coordination}
\end{figure}

The coordinated, uncoordinated, and stasis regimes are identified by Fig. \ref{SIfig:coordination}. Coordinated locomotion is defined by periodic steady states that are synchronous with the actuation cycle, set by $\tau_a$. We measure the state return error as the sum of the squared differences between the segment angles $\theta_i$ after an actuation cycle: 
$$\text{State Return Error} = \sum_i (\theta_i[T] - \theta_i[T + \tau_a])^2.
\label{SIeq:return}$$
In Fig. \ref{SIfig:coordination}a we see a clear boundary between periodic steady states, and multi-period solutions that do not synchronize with actuation. The undulation amplitude has a jump increase at this same boundary, seen in Fig. \ref{SIfig:coordination}b. In the uncoordinated regime the body's stiffness fails to synchronize the motion causing larger amplitude pivots about leg contacts. As the contact-elastic ratio is further decreased and the contact-actuation ratio increased, a third regime is revealed in Fig. \ref{SIfig:coordination}c by a jump in the cost $W^+/d$ (defined in the main text). In this regime, denoted "stasis", locomotion fully jams and the body's motion becomes chaotic. The body is too soft relative to the driving torques, preventing all forward locomotion as segments pivot too far during leg contact to recover synchrony across the body. 

\subsection{Pareto Frontier Analysis}

\begin{figure}
    \centering
    \includegraphics[width=1.0\linewidth]{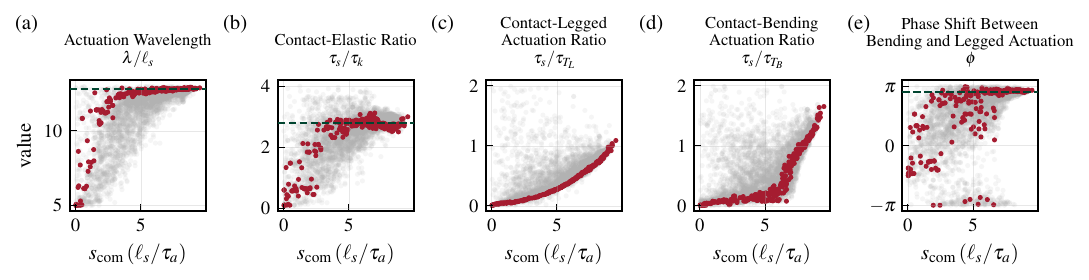}
    \caption{Parameter relationships along the speed-cost Pareto frontier. Red points are on the speed-cost Pareto frontier while grey points are not.  (\textbf{a}) Actuation wavelength increases quickly, reaching just below the maximum of $\lambda/\ell_s=13$. The dashed green line denotes $\lambda/\ell_s = 12.8$, the mean value for high speed solutions. (\textbf{b}) Contact-elastic ratio $\tau_s/\tau_k$ varies for low speeds, but as we transition to more active bending we approach the dashed green line value $\tau_s/\tau_k = 2.8$. (\textbf{c}) Contact-legged actuation ratio increases smoothly with speed. (\textbf{d}) Contact-bending actuation has two regimes. For low speeds, active bending torque is small compared to legged torques. At $s_{com} \approx 6$ we see a slope change, and active bending begins to dominate. (\textbf{e}) At low speeds, $\phi$ varies greatly as it has low influence on the locomotion. As bending increases, $\phi$ approaches a consistent value $\phi=0.9 \pi$, which corresponds to the dashed green line.}
    \label{SIfig:frontier_params}
\end{figure}

After investigating coordination, we turn our attention to the trade off between locomotion speed and efficiency. Given model parameters $\mathbf{x}$, we operationalize this trade off by
\begin{equation}
\label{SIeq:obj}
\mathbf{f}[\mathbf{x}]
=
\begin{pmatrix}
-s_{\mathrm{com}}[\mathbf{x}]\\[2pt]
W^+[\mathbf{x}]/d[\mathbf{x}]
\end{pmatrix},
\end{equation}
where $s_{\text{com}}$ is the body's center of mass speed and $W^+/d$ is the ratio between the positive actuation work and the distance traveled, all measured over an actuation cycle. Varying $\mathbf{x}$ defines a Pareto frontier that characterizes the trade off, which we approximate using MATLAB's multiobjective genetic algorithm \cite{mathworks_gamultiobj}. We consider five free parameters (see Main Sec. Dimensional Analysis and Parameters) with the bounds
\begin{equation}
\label{eq:SI_pareto_bounds}
\mathbf{x}=\Bigl(\lambda/\ell_s,\ \tau_s/\tau_k,\ \tau_s/\tau_{T_L},\ \tau_s/\tau_{T_B},\ \phi\Bigr)\in[5,13]\times[0.01,4]\times[0.01,2]\times[0.01,2]\times[-\pi,\pi].
\end{equation}
For each parameter set $\mathbf{x}$, we simulate the model for $50$ actuation cycles and evaluate Eq. \eqref{SIeq:obj}. We used a population size of $100$ over $50$ generations with \texttt{ParetoFraction}$=0.4$; all other algorithmic options used are MATLAB defaults (including tournament selection and crowding-distance diversity in objective space). Along the estimated Pareto front, solutions did not accumulate at the imposed upper bounds of $\tau_s/\tau_k$, $\tau_s/\tau_{T_L}$, or $\tau_s/\tau_{T_B}$ (SI Fig.~\ref{SIfig:frontier_params}), indicating that optimality is not set by the boundaries chose in Eq. \eqref{eq:SI_pareto_bounds}.

We plot every sampled parameter set $\mathbf{x}$ in Fig. \ref{SIfig:frontier_params} while coloring points on the speed-cost Pareto frontier red. The contact-bending actuation ratio has a knee around $s_{\text{com}}$ corresponding to a transition between leg dominated to bending dominated locomotion. For the high speed bending dominated regime, we find characteristic parameters 
$$
\label{SIeq:typical}
\mathbf{x}[T] = (\lambda/\ell_s = 12.8, \tau_s/\tau_k = 2.8, \tau_s/\tau_{T_L} = T, \tau_s/\tau_{T_B} = 1.45T, \phi = 0.9 \pi),$$
where $T \in [0.5,1].$

\begin{figure}
    \centering
    \includegraphics[width=1.0\linewidth]{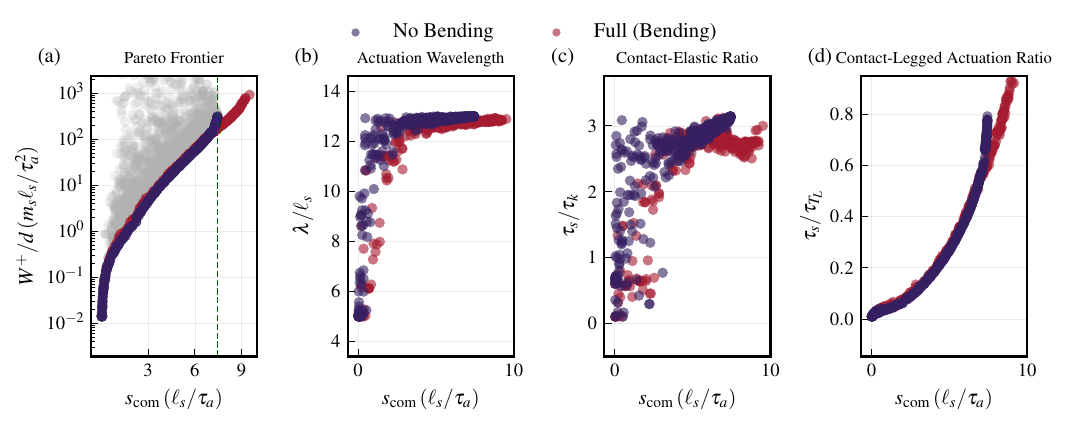}
    \caption{A comparison between the full parameter set Pareto frontier, red points, to a limited frontier that ablates bending by setting $\tau_s/\tau_{T_B} = 0$, purple points. (\textbf{a}) The ablated bending solutions have a similar front for speeds up to $s_{\text{com}} \approx 6$, where the full frontier starts to increase the contact-bending actuation ratio $\tau_s/\tau_{T_B}$ at an increased rate. Without bending, costs diverge reaching a maximum $s_{\text{com}} = 7.45$ denoted by the dashed green line. (\textbf{b}) Both frontiers only use actuation wavelengths that are not close to the maximum $\lambda/\ell_s = 13$ for low speeds. Without bending, this maximum is obtained at high speed, compared to the full frontier which plateaus for the slightly smaller value $\lambda/\ell_s = 12.8$. (\textbf{c}) The contact-elastic ratios are similar with and without bending, until the no bending solution set start to diverge in cost. As this happens the contact-elastic ratio increases, reaching up to $\tau_s/\tau_k = 3.15$. (\textbf{d}) Contact-legged actuation ratios are quite similar for both frontiers, with no bending only deviating at $s_{\text{com}} = 7.25$.}
    \label{SIfig:no_bend_frontier}
\end{figure}

\begin{figure}
    \centering
    \includegraphics[width=1.0\linewidth]{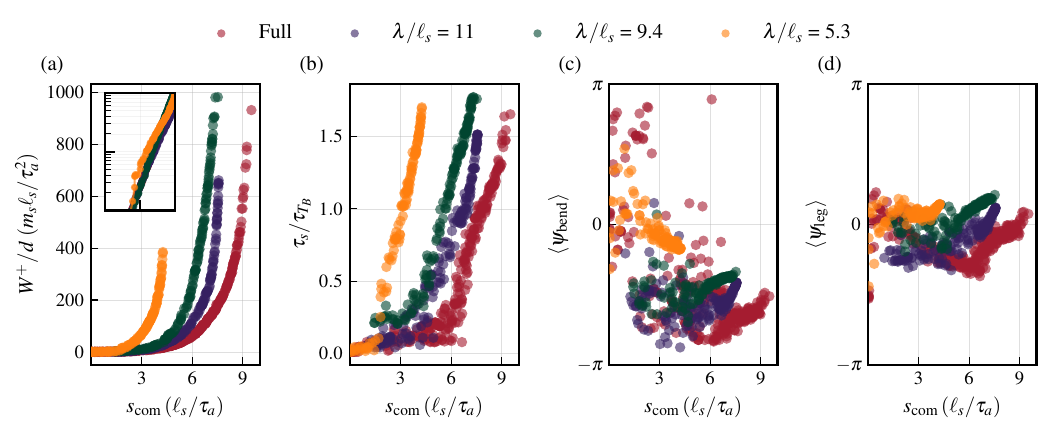}
    \caption{Pareto frontier for fixed actuation wavelength $\lambda/\ell_s$ are similar to the full frontier with variable $\lambda/\ell_s$ up to scaling. (\textbf{a}) The Pareto frontier for fixed $\lambda/\ell_s = (5.3,9.4,11)$, the observed actuation wavelengths in previous experimental work \cite{Anderson1995}. The full Pareto frontier with variable actuation wavelength $\lambda/\ell_s$ is also included. Inset scales each front by their maximal speed and cost with a log $y$-axis, showcasing that the fronts are self similar under scaling. (\textbf{b}) The contact-bending actuation ratio for each frontier, each showing a knee after which bending actuation drives higher speeds. (\textbf{c}) The average $\Psi_{\text{bend}}$ over the body for each parameter set. (\textbf{d}) The average $\Psi_{\text{leg}}$ over the body for each parameter set.}
    \label{SIfig:fixed_lambda_frontiers}
\end{figure}

To better understand this Pareto frontier, we compare the full parameter case with the Pareto frontier found for limited special cases: ablating bending by setting $\tau_s/\tau_{T_B}=0$ in Fig. \ref{SIfig:no_bend_frontier} and fixing $\lambda/\ell_s = [5.3,9.4,11]$ in Fig. \ref{SIfig:fixed_lambda_frontiers}. 

This ablation supports the conclusions of Main Fig. 4, namely that active bending only plays an important role in speed versus cost efficiency for speeds past $s_{\text{com}} = 6$. This is underscored by Fig. \ref{SIfig:no_bend_frontier}a-d, which display similar behavior between the Pareto frontier found with full parameters and with active bending ablated. 

However, these findings do not imply that centipedes or other multilegged terrestrials do not utilize active bending for lower speeds. Speed and efficiency are not the only important objectives for a terrestrial. Stability is also critical, and Fig. \ref{SIfig:no_bend_frontier}b shows that this frontier utilizes near maximal actuation wavelength. Increasing actuation wavelength will have a stability cost, as the average number of legs in contact with the ground throughout an actuation cycle is set by $2N \ell_s/\lambda$. For the maximal $\lambda/\ell_s = 13$ and $N=21$ segments, an average of $3.23$ legs are in ground contact at any time. Lowering actuation wavelength will increase stability, and it is likely that multilegged terrestrials pay the efficiency cost for stability at a given speed.

To quantify how efficiency can be traded for stability, we explore the Pareto frontiers between speed and cost at fixed actuation wavelength values corresponding to experimentally measured values for centipedes at three running speeds \cite{Anderson1995}: $\lambda/\ell_s = 5.3, 9.4,$ and $11$. Each fixed-$\lambda$ frontier exhibits a qualitatively similar speed-cost trade-off to the full frontier, differing primarily in the maximum attainable speed (Fig.~\ref{SIfig:fixed_lambda_frontiers}a). When each front is rescaled by its respective maximum speed and cost (Fig.~\ref{SIfig:fixed_lambda_frontiers}a, inset), the frontiers collapse, indicating an approximate self-similarity in the speed-cost trade-off across stepping wavelengths. Crucially, each fixed-$\lambda$ frontier also exhibits the characteristic knee in the contact-bending actuation ratio (Fig.~\ref{SIfig:fixed_lambda_frontiers}b), confirming that the transition to bending-assisted locomotion is not an artifact of allowing $\lambda/\ell_s$ to vary freely. Notably, these transitions occur at substantially lower speeds for shorter wavelengths, suggesting that active bending may play a role even at moderate walking speeds.

\begin{table}
\centering
\caption{Speed and phase statistics for each Pareto frontier. Model maximum speeds $s_{\mathrm{com}}^{\max}$ are compared with experimental speeds $s_{\mathrm{com}}^{\mathrm{exp}}$ from Anderson et al.\ \cite{Anderson1995}. Phase shifts are reported in units of $\pi$ as mean $\pm$ standard deviation over the top quartile of speed for each frontier, evaluated at segments $k = [5,7,9,11,13]$ as done in Anderson et al. The experimental phase shift $\Psi_{\mathrm{bend}}^{\mathrm{exp}} = -0.10\pi \pm 0.06\pi$ was found to be invariant across speed and longitudinal position \cite{Anderson1995}. See Fig.~\ref{SIfig:fixed_lambda_frontiers}c--d.}
\label{tab:top_quartile_phase_stats}
\begin{tabular}{lcccc}
\hline
Frontier & $s_{\mathrm{com}}^{\max}\;(\ell_s/\tau_a)$ & $s_{\mathrm{com}}^{\mathrm{exp}}\;(\ell_s/\tau_a)$ & $\langle \Psi_{\mathrm{bend}} \rangle/\pi$ & $\langle \Psi_{\mathrm{leg}} \rangle/\pi$ \\
\hline
Full frontier          & $9.51$ & ---                  & $-0.63 \pm 0.05$ & $-0.05 \pm 0.05$ \\
$\lambda/\ell_s = 11$  & $7.57$ & $9.14 \pm 1.26$      & $-0.47 \pm 0.05$ & $0.07 \pm 0.04$ \\
$\lambda/\ell_s = 9.4$ & $7.52$ & $7.35 \pm 0.84$      & $-0.37 \pm 0.01$ & $0.17 \pm 0.01$ \\
$\lambda/\ell_s = 5.3$ & $4.26$ & $5.46 \pm 1.05$      & $-0.16 \pm 0.02$ & $0.12 \pm 0.02$ \\
\hline
\end{tabular}
\end{table}

The maximum speeds obtained at each fixed wavelength and the corresponding experimental values from Anderson et al.\ \cite{Anderson1995} are summarized in Table~\ref{tab:top_quartile_phase_stats}. The model captures the correct trend: the ratio of top speeds between highest and lowest actuation wavelengths is $1.78$ in the model versus $1.67$ experimentally, and the full frontier's maximum speed ($s_{\text{com}} = 9.51\;\ell_s/\tau_a$) falls within the experimental range at the longest wavelength. However, the model systematically underestimates speed at each fixed $\lambda/\ell_s$, as the fixed-$\lambda$ frontiers do not reach the same top-speed as experimentally observed.

This underestimation likely reflects several simplifications in the model. All segments are assigned uniform mass and identical leg lengths, whereas real centipedes exhibit variation in both segment and leg length along the body \cite{Manton1965}. Manton \cite{Manton1965} noted that differential leg lengths in fast-running lineages enable longer effective strides, an effect absent in our uniform model. Active bending is restricted to a single sinusoidal harmonic, precluding the higher-order shape modes that axial musculature can generate. 
Additionally, the free-end boundary conditions on the active bending moment (Eq.~13) impose an asymmetry in which the posterior and anterior segments experience fundamentally different net torques than interior segments, potentially motivating segment-specific control strategies that our spatially uniform sinusoidal actuation does not capture. 
The no-slip contact assumption prevents any foot sliding during stance, eliminating a mechanism by which real legs can redirect ground reaction forces for additional propulsion or maneuverability \cite{Chong2023}. 
Finally, the model's open-loop control lacks the proprioceptive feedback that could allow real centipedes to fine-tune leg timing and body posture within each stride cycle.
Each of these factors could increase the accessible speed at a given wavelength, and their combined absence provides a plausible explanation for the systematic offset between model and experimental speeds.

\subsection{Leg and Bend Phase Shift Analysis}

To understand how body undulations and leg stepping are organized along each fixed-$\lambda$ frontier, we examine the phase relationships between segmental flexion, leg touchdown, and bending activation.
Manton \cite{Manton1965} observed that in fast-running centipedes, leg touchdowns consistently occur on the concave side of each body bend, and argued that this arrangement maximizes the component of leg thrust directed along the body axis. 
In the idealized limit, legs touch down precisely at crests or troughs of lateral undulation, so that at the onset of the pivot the leg force is directed entirely parallel to the direction of motion. 
Passive elastic bending and active muscular bending introduce deviations from this synchrony, and quantifying these deviations provides both a measure of how well the idealized picture captures the actual dynamics and a diagnostic for the extent to which active bending assists or resists the body's undulatory shape change.

We formalize this by defining two phase shifts at each inter-segment joint $k$, where the joint angle is $\sigma_k = \theta_{k+1} - \theta_k$. Both are expressed as $2\pi$ times the normalized cycle-time difference between events. The leg-contact phase shift,
\begin{align}
\label{SIeq:psi_leg}
\Psi_{\mathrm{leg},k} = \frac{2\pi}{\tau_a}\left( \mathrm{argmin}_t [\sigma_k[t]] - t_c^{k+1,R}\right),
\end{align}
measures the time difference between minimal flexion at joint $k$ and the right-leg touchdown at the adjacent anterior segment. The bending-activation phase shift,
\begin{align}
\label{SIeq:psi_bend}
\Psi_{\mathrm{bend},k} = \frac{2\pi}{\tau_a}\left(\mathrm{argmin}_t [\sigma_k[t]] - \mathrm{argmax}_t[\dot{M}[k,t]]\right),
\end{align}
measures the time difference between minimal flexion and the onset of active bending, defined as the zero-crossing of $M_{\mathrm{bend}}$ with positive derivative at joint $k$. Since the flexion $\sigma_k[t]$ is non-linear, with impact events due to leg-touchdown impulses, we measure the minimal flexion time by fitting $\sigma_k[t]$ to a sinusoid. For both phase shifts, a negative value indicates that the reference event (leg touchdown or bending onset) precedes the flexural minimum; a positive value indicates it follows. For comparison with measurements made in Anderson et al.\ \cite{Anderson1995}, we evaluate these quantities at joints corresponding to segments $k = [5, 7, 9, 11, 13]$ in Fig. \ref{SIfig:fixed_lambda_frontiers}c--d. 

Anderson et al.\ reported $\Psi_{\mathrm{bend}} = -0.10\pi \pm 0.06\pi$ rad (adjusting for sign convention), finding no significant variation with either speed or longitudinal position \cite{Anderson1995}. Our model's predictions for the top speed quartile of each frontier are summarized in Table~\ref{tab:top_quartile_phase_stats}. We find qualitative agreement in that bending activation consistently precedes peak flexion ($\Psi_{\mathrm{bend}} < 0$), but the model predicts substantially larger phase leads, particularly at longer wavelengths. Furthermore, in contrast to the experimental finding of speed-invariant phase shifts, our model predicts that $\langle \Psi_{\mathrm{bend}} \rangle$ varies systematically with speed (Fig.~\ref{SIfig:fixed_lambda_frontiers}c). To better understand this discrepancy, we investigate the spatial variation of both phase shifts and the energetic contributions of active bending at each joint.

\begin{figure}
    \centering
    \includegraphics[width=1.0\linewidth]{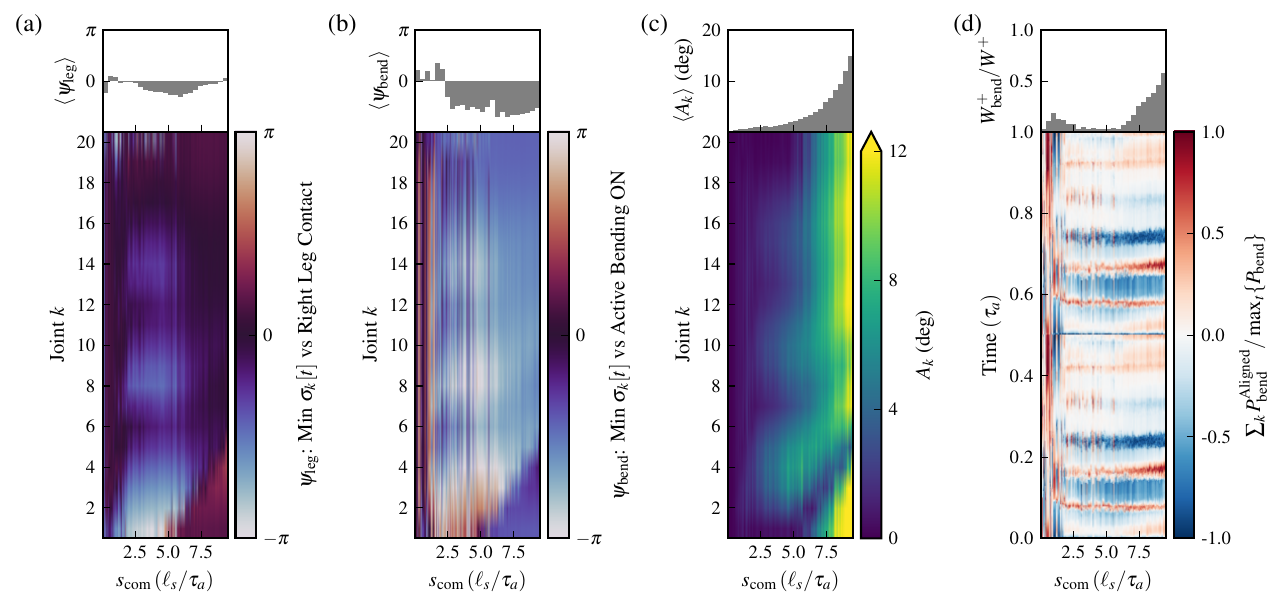}
    \caption{Analysis of active bending for solutions along the cost vs speed Pareto frontier. (\textbf{a}) For each parameter set along the Pareto frontier, we measure the phase shift $\Psi_{\text{leg}}$, as defined in Eq. \eqref{SIeq:psi_leg}.  The top histogram plots the average over the entire body for that speed. A negative lag is introduced as speed increases, with significant variation along the body. This variation diminishes significantly when $s_{\text{com}} > 6 \ell_s/\tau_a$, corresponding to an increase in active bending. (\textbf{b}) Same analysis as (a), but for the phase lag $\Psi_{\text{bend}}$, as defined in Eq. \eqref{SIeq:psi_bend}.  (\textbf{c}) Amplitude of $\sigma_k$ in degrees varies along the body. (\textbf{d}) We sum an aligned power due to bending at each joint, see Eq. \ref{SIeq:p_align} We see clear assistance and resistance provided by bending, which greatly increases in its proportion of total work up to $W^+_{\text{bend}}/W^+ = 0.58.$}
    \label{SIfig:frontier_analysis}
\end{figure}

At the lowest speeds along the Pareto frontier, both $\Psi_{\mathrm{leg}}$ and $\Psi_{\mathrm{bend}}$ exhibit high variability across joints (Fig.~\ref{SIfig:frontier_analysis}a--b, reflecting the broad range of actuation wavelengths $\lambda/\ell_s$ and bending--leg phase shifts $\phi$ sampled at these speeds (Fig.~\ref{SIfig:frontier_params}a,e.
In this regime, active bending contributes little to propulsion, so a specific phase organization is not strongly selected. 

As speed increases and the actuation wavelength converges to $\lambda/\ell_s \approx 12.8$ (Fig.~\ref{SIfig:frontier_params}a), a clear spatial patterning of $\Psi_{\mathrm{leg}}$ and $\Psi_{\text{bend}}$ emerges across the body (Fig.~\ref{SIfig:frontier_analysis}a--b). Since the left and right leg contacts each impose a half-wavelength constraint, the effective synchronization wavelength is $\lambda/(2\ell_s) \approx 6.4$ segments, yielding approximately $N/(\lambda/2\ell_s) \approx 3.3$ wavelengths across the $N = 21$ segment body. The three corresponding minima are visible as bands of negative $\Psi_{\mathrm{leg}}$ and $\Psi_{\text{bend}}$ in Fig.~\ref{SIfig:frontier_analysis}a--b.

The body-averaged leg phase shift $\langle \Psi_{\mathrm{leg}} \rangle$ becomes increasingly negative with speed, reaching a peak magnitude at $s_{\mathrm{com}} \approx 5.91\;\ell_s/\tau_a$. This speed corresponds to the knee in the contact--bending actuation ratio (Fig.~\ref{SIfig:frontier_params}d), marking the transition from leg-dominated to bending-dominated propulsion. Over the intermediate speed range $s_{\mathrm{com}} \in [2.5, 6.0]\;\ell_s/\tau_a$, the body-averaged phase shifts are $\langle \Psi_{\mathrm{leg}} \rangle = -0.23\pi \pm 0.07\pi$ and $\langle \Psi_{\mathrm{bend}} \rangle = -0.52\pi \pm 0.24\pi$ (mean $\pm$ std).

The variation in phase across joints is accompanied by matching variation in flexion amplitudes (Fig.~\ref{SIfig:frontier_analysis}c): joints with larger phase offsets from the ideal $\Psi_{\mathrm{leg}} = 0$ reach higher flexion amplitudes.
Without active bending, this spatial wave of phase decoherence persists, propulsion becomes less efficient, and the maximum attainable speed saturates near $s_{\mathrm{com}} \approx 7.45\;\ell_s/\tau_a$ (Fig.~\ref{SIfig:no_bend_frontier}).
As the system transitions into the bending-dominated regime ($s_{\mathrm{com}} \gtrsim 6\;\ell_s/\tau_a$), active bending reduces the spatial variation in both $\Psi_{\mathrm{leg}}$ and $\Psi_{\mathrm{bend}}$ across the body (Fig.~\ref{SIfig:frontier_analysis}a--b).
In the main text, we identified passive body stiffness as a low-pass filter on the dynamics that enables coordination (Main Fig.~3); here, active bending plays an analogous role in the phase domain.
The bending moment act as a phase-locking mechanism that suppresses the spatial decoherence between leg touchdown timing and body flexion, extending coordinated locomotion to speeds inaccessible through passive stiffness and legged actuation alone.

For $s_{\mathrm{com}} > 6\;\ell_s/\tau_a$, where bending actuation dominates, the anterior joints remain well synchronized with $\Psi_{\mathrm{leg}} \approx 0$, while the posterior joints increasingly transition to a regime with positive $\Psi_{\mathrm{leg}}$ and negative $\Psi_{\mathrm{bend}}$ (Fig.~\ref{SIfig:frontier_analysis}a--b).
This posterior de-synchronization extends progressively toward the body's anterior, affecting the first five segments at the highest speeds.
The corresponding increase in posterior flexion amplitude (Fig.~\ref{SIfig:frontier_analysis}c) indicates that the posterior segments undergo large-amplitude oscillations that are out of phase with the ideal stepping synchrony, consistent with the posterior-driven instability identified in the coordination analysis (Main Fig. ~3).

To understand the energetic role of active bending from the perspective of individual joints, we define the aligned bending power $P^{\mathrm{Aligned}}_{\mathrm{bend}}$ as the instantaneous power dissipated by active bending, summed over all joints, with each joint's contribution time-shifted such that the right-leg touchdown two segments anterior to the joint occurs at $t = 0$:
 \begin{align}
 \label{SIeq:p_align}
P^{\mathrm{Aligned}}_{\mathrm{bend}}[t] = \sum_k P_{\mathrm{bend},k}\!\left[t - \Delta t_k\right],
\end{align}
where $\Delta t_k$ is chosen such that $t_c^{k+2,R} \mapsto 0$ for each joint $k$.
This alignment reveals the temporal structure of active bending's contribution to propulsion relative to the stepping cycle (Fig.~\ref{SIfig:frontier_analysis}d).

At the lowest speeds, $P^{\mathrm{Aligned}}_{\mathrm{bend}}$ is unstructured, reflecting the large variation in step timing set by the actuation wavelength (Fig.~\ref{SIfig:frontier_params}a).
As $\lambda/\ell_s$ converges to ${\sim}12.8$, corresponding to a step duration $\tau_s/\tau_a \approx 0.08$, the aligned power develops clear temporal structure.

Prominent bands of positive (red) and negative (blue) power emerge, corresponding respectively to acceleration and braking at each joint.
The gaps between these bands correspond to the step duration.

The temporal sequence within each half-cycle proceeds as follows.
At $t = 0$, the anterior leg two segments forward the joint touches down; its contribution to the joint is minimal.
When the leg immediately anterior to the joint steps down, a brief high-magnitude acceleration near the impulse event is followed by sustained braking throughout the remainder of that contact.
The subsequent posterior leg touchdown produces an even larger and more sustained acceleration at the impulse, again followed by braking.
Touchdown of the segment two positions posterior to the joint does not repeat this acceleration but instead produces high-magnitude braking at the impulse.
After this sequence, the bending power contribution is small until the cycle repeats for the contralateral side at $t \approx 0.5\;\tau_a$.

For an idealized sinusoidal flexion, the angular velocity $\dot{\sigma}_k$ will change sign concurrent with leg touchdown. 
Given $\Psi_{\text{bend}}$ (Table \ref{tab:top_quartile_phase_stats}) and the leg-bending phase $\phi$ (Fig. \ref{SIfig:frontier_params}) for top speeds along the Pareto frontier, we see that active bending causing acceleration and then deceleration for the anterior and posterior leg touchdowns due to this change in sign. As such, active bending at joints accelerate segments at the start of touchdown until they reach their maximal flexion, and then break for the remainder. This corresponds with the decrease in $\Psi_{\text{leg}}$ when active bending increases (Fig. \ref{SIfig:frontier_analysis}a). 
This is also reflected in Main Fig.~ 4d: the normalized lateral impulse energy loss $\Delta K_\perp / W^+$ increases with speed in the leg-dominated regime, peaking near $s_{\mathrm{com}} \approx 6\;\ell_s/\tau_a$ where it opposes approximately $60\%$ of the positive actuator work.
As active bending increases beyond this transition, $\Delta K_\perp / W^+$ decreases, indicating that bending does not simply add energy to the system but redirects it from lateral to forward motion.
The net effect is that bending's share of total positive work grows to $W^+_{\mathrm{bend}}/W^+ \approx 0.58$ at top speeds (Fig.~\ref{SIfig:frontier_analysis}d), while the energetic cost of lateral losses simultaneously declines.

\begin{figure}
    \centering
    \includegraphics[width=1.0\linewidth]{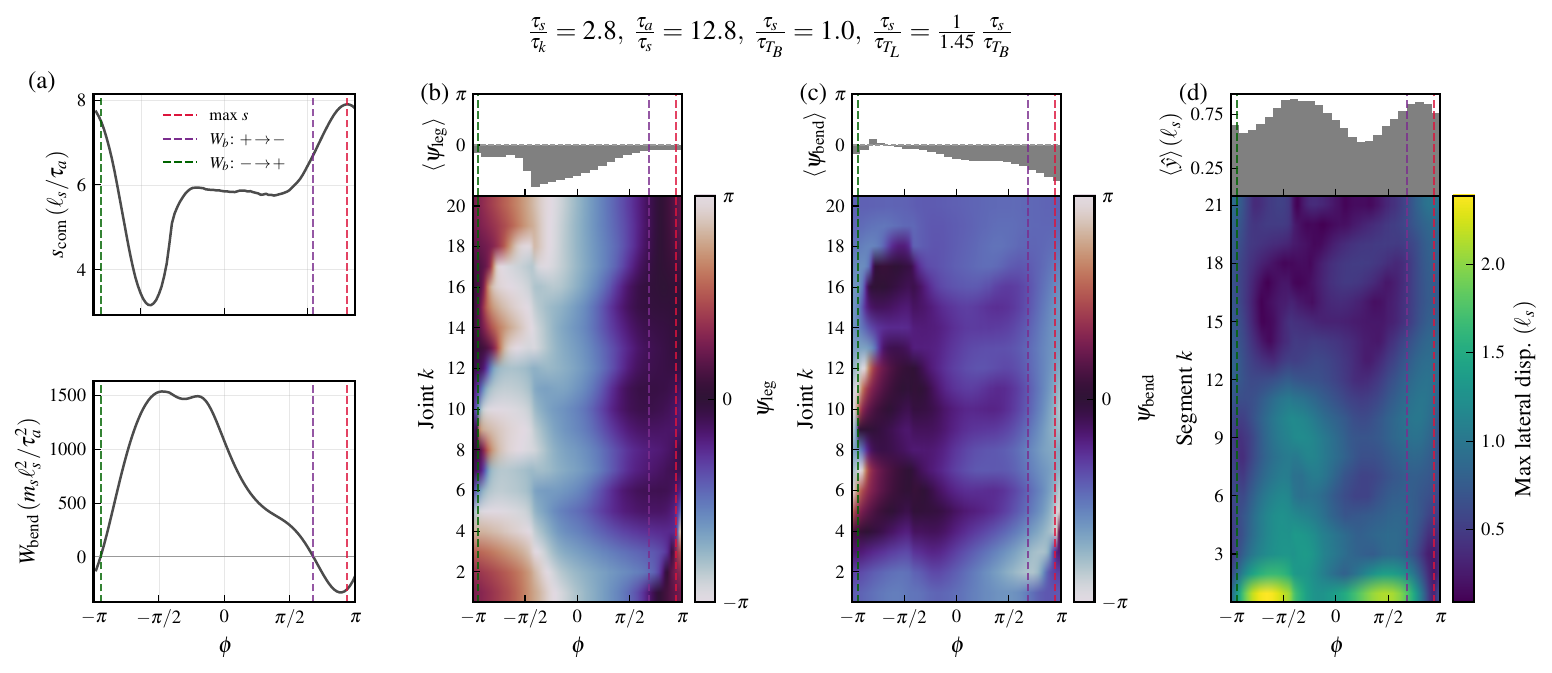}
    \caption{The bending-legged actuation phase shift $\phi$ has a non-trivial role coordinating segments and as a result relationship with $\Psi_{\text{bend}}$. While varying $\phi$, we fix the remaining model parameters by typical values for the high speed regime along the Pareto frontier (see Eq. \eqref{SIeq:typical}).
    (\textbf{a}) As we vary $\phi$, speed varies with a maximum speed corresponding to the phase shift found on the Pareto frontier, $\phi = 0.9 \pi $. Bending work $W_{\text{bend}}$ transitions from positive, to negative, to positive as we vary $\phi$, with dashed lines representing these transitions and also the maximum speed. 
    (\textbf{b}) The phase shift $\Psi_{\text{leg}}$ is minimized near the maximum speed. Increasing or decreasing $\phi$ causes a larger variation in the phase shift between joints. 
    (\textbf{c}) The phase shift $\Psi_{\text{bend}}$ is maximized at the maximum speed.}
    \label{SIfig:sweep_phi}
\end{figure}

To further our understanding of active bending, we investigate how the bending-leg phase shift impacts locomotion for parameters typical to fast speeds on the Pareto-frontier (Fig. \ref{SIfig:frontier_params}). As we sweep $\phi \in [-\pi,\pi]$ we find a maximum speed corresponding to the Pareto frontier value $\phi = 0.9 \pi$. The highest bending work is done at the minimal speed, near $\phi = -\pi/2$ (Fig. \ref{SIfig:frontier_params}a). This corresponds with peak $ \Psi_{\text{leg}}$ and minimal magnitude $\Psi_{\text{bend}}$ (Fig. \ref{SIfig:frontier_params}): although active bending and flexion have synchronized to give the highest bending work, it only occurs while leg touchdowns are phase shifted by almost $-\pi$. 

When $\phi = \pi/2$, corresponding with the optimal assisted bending in the idealized $\Psi_{\text{leg}} = 0$ case, although $\Psi_{\text{leg}}$ has a small negative phase $\Psi_{\text{bend}}$ is considerably negative such that we do not find the required synchrony for large positive bending work. 

Optimal speed is found near $\phi = 0.9\pi$, as predicted by the Pareto frontier, and corresponds to minimal $\Psi_{\text{leg}}$ and maximal $\Psi_{\text{bend}}$. Both phase shifts are relatively homogeneous across joints, and for both increasing beyond $\phi = 0.9\pi$ excites a spatial wave with wavelength set by the actuation wavelength $\lambda/\ell_s$ (Fig. \ref{SIfig:sweep_phi}a--b). The spatial phase dependence can be seen in the maximal lateral displacement, with the maximal lateral displacement having a localized peak in the two posterior segments. We visualize locomotion before and after the peak speed in Movie 3. We see that as we increase $\phi$ past the optimal that the posterior segments increase their oscillation. Notably, this is not the coordination transition seen in Main Fig. 2, as across all $\phi$ the solutions remain stable with a cycle to cycle return error typical for numerical error.

\begin{figure}
    \centering
    \includegraphics[width=1.0\linewidth]{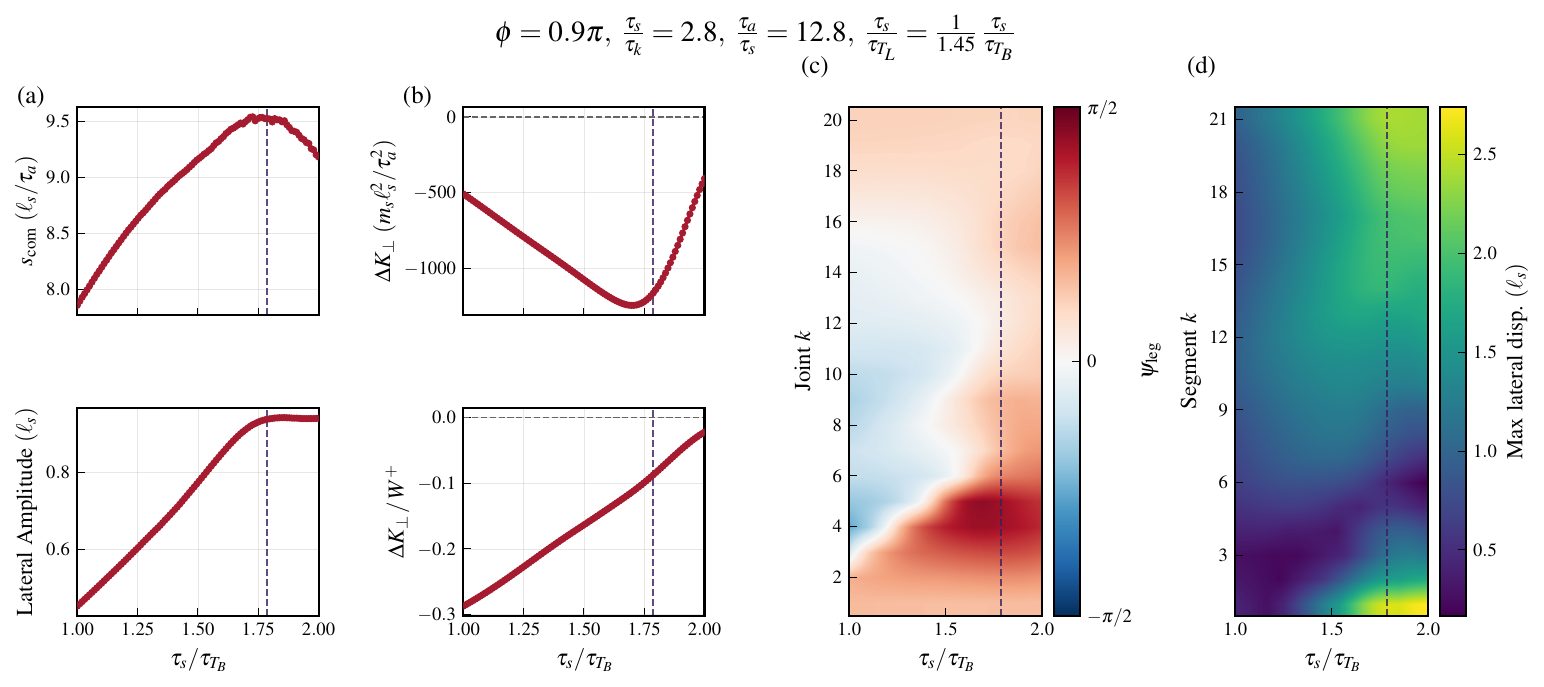}
    \caption{To understand the top speed obtained along the Pareto frontier, we sweep the actuation torque past the fastest speed. we fix the remaining model parameters by typical values for the high speed regime along the Pareto frontier (Eq. \eqref{SIeq:typical}). (\textbf{a}) We reach a maximum speed, indicated by the dashed purple line throughout the figure, at contact-bending actuation ratio $\tau_s/\tau_{T_B} = 1.9$ and contact-legged actuation ratio$\tau_s/\tau_{T_L} = 1.3$. For larger actuation torque, the speed diminishes and the lateral amplitude plateaus. (\textbf{b}) The magnitude of lateral impulse energy loss $\Delta K_{\perp}$ is proportional to the speed, reaching a minimal peak near the maximum speed. The proportion of positive actuation work with $\Delta K_{\perp}$ decreases as actuation torque increases. (\textbf{c}) The flexion-leg phase shift $\Psi_{\text{leg}}$ becomes positive as we increase actuation torque, with high magnitude near the posterior. (\textbf{d}) The maximum lateral displacement achieved during an actuation cycle is localized to the posterior and anterior segments. }
    \label{SIfig:sweep_torque}
\end{figure}

The torque sweep identifies a synchrony-limited performance ceiling. In Fig.~\ref{SIfig:sweep_torque}a, speed increases only up to a finite actuation level and then declines, showing that stronger forcing is not equivalent to better propulsion. The lateral amplitude plateaus when maximum speed is reached. 

Counter-intuitively, when we start to lose speed the lateral kinetic energy lost due to impulses, $\Delta K_{\perp}$. decreases in magnitude (Fig. \ref{SIfig:sweep_torque}b).  The proportion of positive actuation work with $\Delta K_{\perp}$ does decrease, suggesting that past the top speed actuator work drives shape change rather than translational motion. 
This is supported by both Fig. \ref{SIfig:sweep_torque}c--d which show that the phase shift between peak flexion and leg touchdowns becomes positive and the boundary joints increase in maximum lateral displacement. Visualized in Movie 4, we see that higher actuation torque leads to large amplitude oscillation in the posterior. 

Fig. \ref{SIfig:frontier_analysis} supports a breakdown in synchrony. When $\Psi_{\text{leg}}$ is negative, active bending accelerates segments at the start of touchdown such that legs can deliver efficient forward propulsion. But as we increase the actuation torque, active bending accelerates segments too far such that $\Psi_{\text{leg}}$ becomes positive. After this, active bending does net positive work (Fig. Main 4b) and further increase in torque will cause greater rotation that induces inefficient shape changes that prevent legs from driving the body forward.

\end{document}